# Chemical vapour deposition synthetic diamond: materials, technology and applications

R S Balmer, J R Brandon, S L Clewes, H K Dhillon, J M Dodson, I Friel,
P N Inglis, T D Madgwick, M L Markham, T P Mollart, N Perkins, G A
Scarsbrook, D J Twitchen, A J Whitehead, J J Wilman and S M Woollard

Element Six Ltd., Kings Ride Park, Ascot, SL5 8BP, UK
Corresponding authors: Joe.Dodson@e6.com, Daniel.Twitchen@e6.com

**Abstract.** Substantial developments have been achieved in the synthesis of chemical vapour deposition (CVD) diamond in recent years, providing engineers and designers with access to a large range of new diamond materials. CVD diamond has a number of outstanding material properties that can enable exceptional performance in applications as diverse as medical diagnostics, water treatment, radiation detection, high power electronics, consumer audio, magnetometry and novel lasers. Often the material is synthesized in planar form, however non-planar geometries are also possible and enable a number of key applications. This article reviews the material properties and characteristics of single crystal and polycrystalline CVD diamond, and how these can be utilized, focusing particularly on optics, electronics and electrochemistry. It also summarizes how CVD diamond can be tailored for specific applications, based on the ability to synthesize a consistent and engineered high performance product.

**PACS:** 81.05.Uw, 03.67.-a, 87.64k, 81.05.Cy

## 1. Introduction

Historically, diamond used in industrial applications has been appreciated for its combination of extreme hardness and wear resistance. However, it has long been recognized that diamond is a remarkable material with other properties – optical, thermal, electrochemical, chemical, electronic – that also outclass competing materials. When these properties are combined, they offer the designer an engineering material with tremendous potential, offering solutions that can shift performance to new levels or enabling completely new approaches to challenging problems.

Diamond grown using chemical vapour deposition (CVD) processes has released the broad engineering potential of diamond by making available material with consistent characteristics, tailored to end applications. This is in contrast to the wide variation in natural diamond, where the challenges of selecting individual stones with the right characteristics and properties for a particular engineering application put a prohibitive cost on its use in all but a few niche areas (figure 1).

In 1992 there were a few high technology application areas of diamond, including optoelectronic switches, radiation detectors, transmission components and lasers, alongside the more traditional uses such as anvils, hardness indenters and styli for materials science applications, surgical blades and heat sinks. Most or all of the diamond products sold for high technology applications were made from either natural or High Pressure, High Temperature (HPHT) synthetic diamond (section 2.1). CVD diamond had only been used as an X-ray mask support and was expected to be used as an infra-red transmission window within a few years (Seal 1992). The first commercial CVD products were being produced for abrasive applications in the machining sector, and for heat spreading.

In 2009, many grades of synthetic diamond are available from several suppliers, with more routine materials being available via e-shops with the click of a mouse and a credit card (www.e6cvd.com). Polycrystalline diamond windows in excess of 100 mm diameter and 1.5 mm thick are routinely fabricated for optical and microwave transmission windows (Godfried *et al.* 2000, Thumm 2007) and thick free-standing single crystal CVD diamond is fabricated into optical components with at least one dimension in excess of 5 mm (Mildren *et al.* 2008) or with such exceptional purity as to make the material suitable for a number of electronic applications (Isberg *et al.* 2002, Twitchen *et al.* 2004, Secroun *et al.* 2007, Nesladek *et al.* 2008).

Synthesis of diamond by CVD has not only contributed to reducing the costs of the material but has enabled the extreme properties of diamond to be fully utilized. Components using CVD diamond now span tweeters for loudspeakers, radiation detectors and sensors, optical components for lasers, windows for RF and microwave transmission, a range of blades and cutting tools, and electrodes for electrochemical sensing, ozone generation and direct oxidation of organic matter. Further, in active development there are new applications including high power electronics, quantum optics and erosion resistant coatings in nuclear fusion reactors.

Developments in material synthesis and processing are leading to high purity and low defect grades of the material, such as an ultra-low birefringence single crystal CVD diamond (Friel *et al.* 2009). This provides a step forward for all photonics applications of diamond, and is being trialled in laser applications. Quantum purity grade diamond, in which both chemical and isotopic impurities are controlled (Balasubramanian *et al* 2009), complements electronic, optical and thermal grades already in commercial production.

Over the last few years several companies have been established to exploit the commercial potential of CVD diamond material. These cover wastewater treatment systems; ozone generation equipment; developing active electronic devices using CVD diamond; and diamond detectors for markets including medical X-ray dosimetry, radiation monitoring in the nuclear and defence industries, data logging in oil well exploration and UV photolithography for semiconductor manufacturing.

These investments, coupled with the broad range of applications being developed elsewhere, show that a new age of industrial diamond has truly begun – that of CVD diamond.

## 1.1 Scope of this review
This paper provides an up to date overview of the key properties of diamond in its main applications areas. The origin of those properties and their sensitivity to defects in the diamond is explained, so that the engineer can more easily understand the appropriate specification for the diamond, and the main design considerations when using diamond. This review focuses on applications of diamond based on its bulk material properties: mechanical, thermal, optical, electronic and electrochemical, and does not include areas such as: quantum applications (Jelesko and Wrachtrup 2008), biological applications of diamond (Nebel *et al.* 2007a,b) or nanodiamond (grown by CVD (Zhou *et al.* 1998) or separated by milling (Liang *et al.* 2009)) which has demonstrated a range of applications (Rittweger *et al.* 2009, Williams *et al.* 2008, Hu *et al.* 2008, Auciello *et al.* 2004).

## 2. Growth of diamond

### 2.1 Thermodynamically stable synthesis - HPHT and natural diamond

At room temperature and pressure, graphite is the stable allotrope of carbon while diamond is a metastable allotrope. The genesis of natural diamond is believed to occur at depths of around 200 km, corresponding to pressures and temperatures of 70-80 kbar and 1400-1600°C (Blatt and Tracy 1996, Finnerty and Boyd 1984). This is in the region of the carbon phase diagram where diamond is thermodynamically stable (figure 2).

Natural diamonds (figure 1) have grown in a variety of uncontrolled environments and their composition and growth habits vary significantly. Robertson *et al.* (1934) divided them into types I and II on the basis of optical absorption. We now know that the extra absorption between 230 and 300 nm and in the infra-red around 8 µm in type I diamonds is due to the presence of nitrogen. Other properties are summarized in table 5.

The first industrialized method of synthesising diamond was the HPHT method (Bundy *et al.* 1955). Diamond is produced in the thermodynamically stable regime from a locally carbon-rich melt. In contrast to natural genesis, the presence of a solvent metal, such as Fe, Co or Ni, was found to be essential to a practical process (Bundy 1963). Pal'yanov *et al.* (1997) presents a review of HPHT growth.

The HPHT growth process offers a significant degree of control over the quality and geometry of diamond obtained. Most diamond produced by this method is small grains of type Ib for use in grinding and other abrasive applications. Longer, controlled growth periods enable production of single crystal Ib stones with dimensions routinely up to 8 mm, again typically used in abrasive tools. Section 3.2 discusses the wear properties of diamond.

Addition of selected getter mixtures (for example Ti, Al, Zr) to the capsule enables growth of type IIa diamond (Burns *et al.* 1999, Sumiya *et al.* 2002). In this case any nitrogen present is preferentially bound to the getter, rather than being incorporated in the diamond lattice. Extraordinary degrees of lattice perfection have been achieved in processes similar to this, where the slow, well-controlled growth has generated areas of over $4 \times 4 \text{ mm}^2$ with no detectable extended defects (Burns *et al.* 2007).

### 2.2 Metastable synthesis - CVD

Diamond can also be synthesized in its metastable regime. Growth by Chemical Vapour Deposition (CVD) was first performed in 1952 and work leading to industrialized processes was reported in the 1980s (Martineau *et al.* 2004 and references therein). More recently, synthesis under a variety of hydrothermal conditions has been reported (Chen and Chen 2008, Szymanski *et al.* 1995, Zhao *et al.* 1997, Roy *et al.* 1996, Korablov *et al.* 2006) although the technique has not been widely developed.

Being in the region where diamond is metastable compared to graphite, synthesis of diamond under CVD conditions is driven by kinetics and not thermodynamics. Diamond synthesis by CVD is normally performed using a small fraction of carbon (typically <5%) in an excess of hydrogen. If molecular hydrogen is heated to temperatures in excess of 2000K, there is a significant dissociation to atomic hydrogen. The heating can be by arc-jet (Luque *et al.* 1998, Owano *et al.* 1991, hot filament (Herlinger 2006), microwave plasma (Sevillano 1998) or DC arc (Abe *et al.* 2006, Baik *et al.* 1999) or even by using an oxy-acetylene flame (Bachmann *et al.* 1991). In the presence of a suitable substrate material (table 1), diamond can be deposited.

With a well designed reaction chamber, the majority of the impurities in the gas mixture come from the source gases. Many gases are now available commercially containing less than 1 ppm total impurities. The low concentrations of source gas impurities have contributed to the growth of diamond with exceptionally high purity (Twitchen *et al.* 2004, Tallaire *et al.* 2006, Kasu and Kobayashi 2003) applicable to electronic and quantum applications.

As well as controllable purity, another significant aspect of CVD diamond growth is that large areas of diamond can be obtained. While very few natural, single crystal gem diamonds have dimensions exceeding 15 mm, freestanding polycrystalline diamond wafers are routinely manufactured in discs exceeding 100 mm (Baik *et al.* 1999, Heidinger *et al.* 2002, Parshin *et al.* 2004). Thin diamond coatings are available with dimensions exceeding 300 mm (Herlinger 2006).

In contrast to HPHT or natural diamond, CVD synthesized diamond can also be deposited conformally on shaped substrates, enabling growth of items such as speaker domes (section 3.3), shaped and patterned components (Wörner *et al.* 2001, Ribbing *et al.* 2003, Jubber *et al.* 1998) and missile domes (Wort *et al.* 1999).

*2.2.1 Forms of CVD diamond.* Most macroscopic solid materials can be divided into three broad classes - amorphous, polycrystalline and single crystal. Amorphous materials are characterized by a lack of long range order. They include glasses, plastics and diamond-like carbons (DLCs). Polycrystalline materials are also common, including most metals, metal alloys, and igneous rocks. These are composed of small single crystals (grains) bound tightly together by a thin disordered interface.

Macroscopic single crystals do exist, but are much less common. They can be found in nature, including gem diamonds, or as occasional rock crystals tens of centimetres long. Engineered single crystals of metal superalloys are used in jet engine turbine blades. Some optical devices, including lasers, use single crystals including sapphire. The most widely used macroscopic single crystal of our era is silicon. Semiconductor grade silicon is grown in boules up to 400 mm diameter and over 1 metre long, before being extensively sliced, diced and processed into the ubiquitous electronic chip.

Diamond grown by CVD can fall into two of these categories - polycrystalline and single crystal. Single crystal diamond is formed by homoepitaxial growth, where the seed, or substrate, is a single crystal diamond (natural, HPHT, or CVD). Polycrystalline diamond is typically formed whenever growth occurs on a non-diamond substrate.

Unlike many other polycrystalline materials, the grain structure of polycrystalline diamond has a non-uniform composition. The grains show a preferred direction, and vary dramatically in size from one side of a polycrystalline diamond wafer to the other. This is due to the growth process. A high density of nucleation sites (up to $10^{12}$ cm$^{-2}$ in Baldwin *et al.* (2006)) is initially formed on the substrate. Each of these sites grows, but grains with favoured facets and orientations grow preferentially with respect to their less favoured neighbours. This growth competition increases the size of grains to around 10% of the layer thickness; each surviving grain can be traced back to a single nucleation point at the substrate surface.

In addition to homoepitaxial growth, significant efforts have gone into single crystal diamond growth by heteroepitaxy. The non-diamond substrate is chosen so that nucleating grains are crystallographically aligned, and strain and lattice mismatch are minimized. Careful control of the growth conditions is required to ensure that the grains remain aligned in both the growth plane and growth direction and can coalesce. Films up to 100 mm diameter have been demonstrated on iridium with a maximum thickness of around 50 µm (Fischer *et al.* 2008). As with heteroepitaxial growth of other materials, extended defects propagate from the original grain boundaries but, unlike them, diamond cannot be readily annealed to create a low defect single crystal.

*2.2.2 CVD growth conditions.* The features uniting most methods of growing diamond by CVD are gas temperatures in excess of 2000 K, and a gas phase carbon fraction of a few percent in a background of hydrogen. Common variations include oxygen addition and addition of up to a few percent of nitrogen, as well as elements added explicitly for doping

The carbon fraction can be supplied in many forms - if simple hydrocarbons are used, it has been shown that there is no significant difference between the results obtained using $CH_4$ or $C_2H_2$ as source (Cheesman *et al.* 2006) as there is rapid interconversion between $C_1$ and $C_n$ species.

Addition of oxygen has been suggested to enable deposition at lower substrate temperatures, be beneficial to the quality of the diamond under some circumstances, but may be responsible for a weakening of polycrystalline diamond (Petherbridge *et al.* 2001, Yan *et al.* 2002, Harris *et al.* 1999). The tie-line dividing diamond growth and etching in the Bachmann diagram (Bachmann *et al.* 1991) approximately links CO and $H_2$. These can be considered to be two "background" gases for CVD diamond growth, with CO being formed whichever combination of oxygen and carbon source gases is used.

Nitrogen is difficult to eliminate from the growth of diamond by HPHT and is incorporated in different amounts in different growth sectors. Its near ubiquity in natural diamonds demonstrates that it is often involved in natural genesis. For growth by CVD the addition of ppm levels of nitrogen has been shown to modify the morphology and texture of polycrystalline films, increase the growth rate (Locher *et al.* 1994) and eventually weaken the grain boundaries. An extreme example of its involvement is in growth of single crystal CVD diamond at substrate temperatures ~1100°C. In this case, its presence is essential at levels between 0.1% and 0.6% (Yan *et al.* 2002) and may be playing a role in surface stabilisation as well as its normal role in catalytic growth rate enhancement.

While some polycrystalline diamond can be grown under most conditions in table 1, few conditions will produce high quality growth on all facets of diamond. Morphology models can predict development of the shape of single crystal diamond (Wild *et al.* 1993, Silva *et al.* 2006). If the {111} faces in figure 3 do not grow epitaxially, possibly due to twinning (Sawada and Ichinose 2005), this would lead to termination, rather than growth, of the single crystal. However, the same growth conditions may be acceptable for growth of polycrystalline diamond with its initial random orientation of grains.

**Table 1.** Summary of conditions which can lead to diamond growth if the gas temperature is in excess of 2000K, such as in a microwave plasma. [The hot plasma conditions mean that there is rapid interconversion between $C_1$ and $C_2$ species such that replacing $CH_4$ with $C_2H_2$ etc leads to similar results.]

| Condition | Acceptable range |
| --- | --- |
| Substrate temperature | 700-1300°C |
| Possible substrate materials | Si, Mo, W, Ti, diamond, materials with low carbon solubility |
| $\dfrac{[CH_4]}{[H_2]+[CO]}$ | 0.1-10% |
| $\dfrac{[N_2]}{[CH_4]+[H_2]+[CO]}$ | 0-0.6% |

While polycrystalline diamond can be grown over large areas, similar sizes of single crystal diamond substrates are not available. HPHT diamond has been synthesized up to approximately 20 mm in one dimension (Koivula *et al.* 1993), but is ultimately limited by the size of capsule used and stability of the process for a long run. Progress has been made in extending the size of diamond through CVD synthesis both laterally (Silva *et al.* 2008) and vertically up to around 10 mm (Mokuno *et al.* 2006); single crystal CVD synthetic diamond plates up to 8 mm square are available commercially (www.e6cvd.com). Ways of joining multiple diamond single crystals together to create a single larger area crystal has also been the subject of patent applications (e.g. Meguro *et al.* 2005), but this generally suffers many of the problems of heteroepitaxy, and is not thought to be in widespread use.

### 2.3  Defects in diamond grown by CVD
Perfect diamond would consist entirely of $sp^3$ bonded carbon at every point on the lattice. In polycrystalline diamond, the most obvious deviation from this is the grain boundaries. While forming the boundary between crystallites with different orientations, they consist largely of disordered $sp^3$ bonded carbon, and a high proportion of hydrogen (Reichart *et al.* 2003), presumably at dangling bonds. However, as long as the bulk properties remain dominated by the grains and not the grain boundaries, polycrystalline diamond can be a technologically excellent product showing similar properties to that of bulk single crystals.

Once the grain boundaries are removed, and single crystal diamond is investigated, extended and point defects become more apparent. Extended defects include twins, vacancy clusters (Hounsome *et al.* 2006) and dislocations. The effects of dislocations in single crystal CVD diamond are discussed in section 5.6.

Moving to smaller scales, we arrive at point defects. A great variety of point defects occur in diamond (optically active absorption by defects take up 247 pages in Zaitsev (2001)). One class of point defects that is rarely reported is the interstitial, as the high atomic density and rigid lattice leave little room for foreign atoms. Despite the high atomic density, small substitutional defects are found to occur, such as nitrogen and boron, presumably due to non-equilibrium chemical bonding processes occurring on the diamond growth surface. These can act as electron donors and acceptors (section 6.1). Other non-carbon atoms that have been identified in CVD diamond in defect complexes include Si, P and S (Edmonds *et al.* 2008, Zaitsev 2001, Gheeraert *et al.* 2002).

Many complexes of defects including vacancies and hydrogen have been identified and studied by techniques including absorption, luminescence and EPR spectroscopies (Glover *et al.* 2003, Glover *et al.* 2004, Baker 2007, Felton *et al.* 2008). The ability to control the isotopes in the CVD growth environment has assisted in discovering the nature of many defects. A further degree of freedom in defect complexes is charge. For example, Fritsch *et al.* (2007) have suggested that charge transfer between defects is the cause of colour changes in natural "chameleon" diamonds, and Khan *et al.* (2009) (this issue) reports similar effects in CVD diamond.

Many defects in CVD diamond are associated with $sp^2$ bonded carbon, which can be identified in Raman spectroscopy. The sensitivity of Raman spectroscopy to $sp^2$ bonding increases with excitation wavelength (Wagner *et al.* 1991, Prawer and Nemanich 2004). In high quality diamond, the peak attributed to $sp^2$ carbon can be barely detectable even with infra-red excitation.

*2.4 Different grades in diamond*

The properties of CVD diamond are strongly influenced by the grain structure and impurity content (Morelli *et al.* 1991, Koidl and Klages 1992). Simply, larger grained higher purity material results in higher thermal conductivity and lower optical absorption, while finer grained layers have superior mechanical performance. These parameters can be controlled by the synthesis conditions, thus the characteristics of CVD diamond films can be tailored with different properties as illustrated in figure 4.

As the applications for polycrystalline CVD diamond increase, there is constant development and refinement of the properties. Large area polycrystalline diamond plates are used in both detector and RF window applications and specific grades have been developed with optimal characteristics for these applications (Whitehead *et al.* 2001, Brandon *et al.* 2001). As new applications and manufacturers emerge no doubt different optimized grades of CVD diamond will follow.

Single crystal CVD diamond grown homoepitaxially on HPHT substrates has followed a similar pattern, becoming a commercially available material earlier this decade. Grades optimized for cutting, electronic and optical applications are emerging.

## 3. Mechanical and wear properties

*3.1 Diamond's strength*

The strength of diamond is attributable to its high density of very strong bonds. Failure is typically by crack propagation, and dislocation creep does not occur in diamond at normal temperatures. Field (1992) extensively reviewed the literature relating to the strength of natural diamond; it is apparent that there are practical difficulties in measurement because of the extremely limited availability of diamond in large sizes and the high cost of obtaining and processing such material.

A theoretical analysis based on the elastic modulus, fracture surface energy and nearest neighbour distance gives a theoretical strength of about 200 GPa (~one fifth of Young's modulus). Applying Griffith's equation with crack lengths of order 10 nm gives a theoretical strength approaching 10 GPa (Field 1992).

Clearly diamond is a potentially very strong material, but its behaviour is highly notch sensitive (meaning that stress concentrating surface flaws have a large effect on the fracture strength due to the extremely low level of plastic deformation at the tip of the flaw) and therefore its measured strength is not single-valued and has to be treated using Weibull statistics (for example Waterman and Ashby (1991)). Thus, the strength of a diamond is dependent upon its history and particularly the last stage of surface processing to which it has been subjected.

*3.1.1 Single Crystal Diamond.* Measuring the strength of single crystal diamond is technically challenging because cost necessitates the use of small samples. Much work has used indentation methods as the volume of material required is much lower (see Brookes (1992) for a thorough review of this data). The consistent preparation of sample surfaces for mechanical tests is a major problem and almost certainly affects the results and makes their reproduction difficult.

As part of an unpublished study (Whitehead 1995), a jig was specially made for clamping the end of a beam. The clamp was designed to clamp parallel and the clamping surfaces were made from polycrystalline cubic boron nitride to minimize the extent of deformation at the clamped edge. Plates about 5 mm x 3 mm were prepared from large HPHT and natural type IIa single crystals, with all surfaces polished to an $R_a$ (mean modulus surface deviation) of less than 1 nm. 15 HPHT plates (thickness 0.5 mm) and five natural plates (thickness 0.2 mm) were tested. The results are in table 2, although no Weibull modulus is given for the natural samples. In all cases, the failure mode was cleavage on one or more {111} planes.

A later study using the same testing protocol (Houwman 2003) measured the strength of a large number of CVD single crystal plates together with a further 14 natural type IIa samples. All of these samples were polished to a surface roughnesses ($R_q$ – root mean square surface deviation) of order 1 nm. Strengths of up to 5.1 GPa were measured for the CVD single crystal diamond plates and around 4.5 GPa for the type IIa natural diamond plates. However the spread of values obtained was wide and the Weibull modulus was low. Applying the flaw size model in Field (1992), the strength-limiting flaw size is between 0.1 μm and 5 μm for the maximum and minimum measured strengths.

**Table 2:** Strength of single crystal diamond measured using cantilever beam test.

| Diamond | Strength, MPa | | Weibull Modulus, ± 1 standard error |
| --- | --- | --- | --- |
| | Mean | Sample Standard Deviation | |
| HPHT synthetic (15 samples) | 1500 | 325 | 5.2 ± 0.2 |
| Natural type IIa (5 samples) | 2350 | 400 | Not calculated |
| Natural type IIa (14 samples) | 2790 | 1050 | 2.48 ± 0.10 |
| CVD synthetic (30 samples) | 2860 | 1200 | 2.55 ± 0.12 |

It is clear that whilst single crystal diamond can have a very high strength, its strength can be degraded by the presence of flaws. The most likely source of flaws is believed to be from surface processing and subsequent handling of the material.

To date there have been no systematic studies of the strength of single crystal diamond as a function of nitrogen content.

*3.1.2 Polycrystalline Diamond.* Suitably sized material is more readily available to perform strength measurements on polycrystalline CVD diamond, but interpretation of the results remains complex.

Early measurements were made by subjecting thin membranes (between a few μm and about 100 μm or so), to differential pressure with compressed gas and measuring the bursting stress (e.g. Cardinale and Robinson 1992). For films of comparable thickness, higher strengths were observed with the nucleation surface in tension than with the growth surface in tension. Valentine et al (1994) performed similar tests using diamond discs 200-300 μm thick and found that for damond from the same synthesis run (and therefore produced presumably under the same conditions), higher strengths were measured with the nucleation surface in tension.

A more wide-ranging study, which involved breaking several hundred samples using a three-point bending geometry, has enabled the strength of a 'mechanical' grade and an 'optical' grade of polycrystalline CVD diamond to be evaluated as a function of thickness and a model developed for the strength-limiting flaws in the material (Pickles 2002). The strength of polycrystalline diamond is limited by flaws that are typically the lateral size of the grains at the tensile surface; flaws introduced by processing do not appear to play a significant role. The strength is found to decrease with thickness with either the growth or nucleation surface in tension. However, the nucleation surface strength is largely independent of the synthesis process, whereas there is a measurable difference for the growth surface. This is attributed to the different rates at which the grain size at the growth surface evolves with thickness. One consequence of the strength limiting flaws being intrinsic to the material is that the strength of polycrystalline CVD diamond produced by a particular process is reasonably predictable, particularly when the growth surface is in tension. The range of strengths observed in high quality polycrystalline diamond is summarized in table 3.

**Table 3:** Strength of nominally 500 µm thick polycrystalline diamond measured using 3-point bend test (Pickles *et al.* 1999, Pickles 2002).

| Diamond | Strength, MPa | | Weibull Modulus |
|---|---|---|---|
| | Mean | Sample Standard Deviation | |
| Nucleation side (all grades) | 1151 | 236 | 6.5 ± 0.2 (optical) |
| | | | 11 ± 0.5 (mechanical) |
| Growth side (optical grade) | 557 | 75 | 23.1 ± 0.5 |
| Growth side (mechanical grade) | 735 | 150 | 11.3 ± 0.5 |

*3.2 Diamond's wear properties*

The high strength of the carbon to carbon bond is the source of the exceptional mechanical and wear properties of diamond. Dislocation of the atoms is difficult and consequently diamond is the hardest known substance on any measurable scale (table 4). Combining its wear properties with low friction and high thermal conductivity makes diamond a very desirable material for cutting tool applications. Its high speed particle erosion resistance has been studied in detail by Davies and Field (2002) and Kennedy *et al.* (1999).

**Table 4.** Hardness properties of diamond (Field 1992).

| Property | Value (GPa) |
|---|---|
| Knoop hardness (500 g load) | {001} surface = 56 − 102<br>{111} surface = 88 |
| Vickers Hardness | Polycrystalline = 85 − 100<br>Single Crystal = 70 -100 |

In the authors' wear tests, turning on high silicon-aluminium alloys showed that the wear properties vary for different types of diamond and also as a function of the primary cutting orientation. Tools made using a {110} orientation showed lower wear than those with a {100} plane.

Comparing tools with a {110} orientation, it was found that SC CVD has around 20% and 45% higher chip resistance respectively than natural type IIa and HPHT type Ib diamond. Abrasion resistance, in the steady state, was highest for {110} CVD and HPHT, with the CVD tools exhibiting roughly 60% higher wear resistance than natural type IIa.

It is not entirely clear why tools fabricated with {110} orientation are more wear resistant than those fabricated from a {100} plane (figure 4), but it is thought to be associated with two factors:

(i)      The ability to prepare a fine surface finish, free of damage which can promote crack initiation, can vary from one diamond surface to another

(ii)      The angle the cutting face makes with the hardest diamond wear direction <111>.

While natural diamond samples can also show excellent performance, the variability in starting material (table 5) is such that it can be very difficult to predict the exact performance, and no two natural tools behave the same.

**Table 5.** Comparison of properties for natural and synthetic diamond.

| Property (typical values) | Type IIa SC CVD | Type Ib HPHT | Type Ia Natural | Type IIa Natural |
|---|---|---|---|---|
| Nitrogen Content (ppm) | <1 | ~150-200 {100} sectors | 200-3000 | <10 |
| Dominant nitrogen form | Single substitutional | Single substitutional | Aggregates | Aggregates or substitutional |
| Colour (1 mm plate) | Colourless | Yellow | Colourless → yellow to brown tint | Brown/ colourless |
| Thermal Conductivity $(Wm^{-1}K^{-1})$ (300K) | 1800-2200 | ~800-1200 | 400-1200 | 1800-2200 |
| Stress $\Delta n$ (1 mm plate) | $<5x10^{-5}$ | $<1x10^{-5}$ | *Varies from* $1x10^{-6}$ *to* $1x10^{-2}$ | *Varies from* $10^{-4}$ *to* $10^{-2}$ |
| Dislocations | $<10^{4}$ $cm^{-2}$ | $10^{4}$ -$10^{6}$ $cm^{-2}$ | $<10^{6}$ $cm^{-2}$ | $10^{8}$ -$10^{9}$ $cm^{-2}$ |

*3.2.1  Examples of diamond's use in wear and cutting applications.*  Over recent years the machining industry has seen an increasing demand for working with highly abrasive nonferrous material and metal matrix composites, commonly used in the automotive and aerospace industries. Diamond is an ideal material for machining these materials to the high tolerances and surface finishes required. In addition diamond's high thermal conductivity and excellent thermal oxidation resistance makes it suitable for both high speed and dry machining. Diamond cutting tool applications also include dressers for wheel dressing, speciality knives, burnishing tools, and wire drawing dies. CVD diamond can also be engineered to make it electrically conductive and therefore making it cuttable using electrical discharge machining (EDM) (Collins 1998) so that tailored geometries are more easily attainable.

CVD diamond's performance is increasing the range of industrial cutting applications diamond can address.  For example the lifetime of diamond water jet nozzles used for cutting nappies, cloth, paper, cardboard, frozen foods, semiconductors etc can be >100 hrs compared to tens of hours for more standard ruby or sapphire nozzles.

*3.3 Diamond used in acoustic applications*

Current commercial applications making use of the high stiffness of diamond include micromechanical oscillators (Baldwin *et al.* 2006), surface acoustic wave (SAW) devices (Shikata 1998) and tweeter domes for loudspeakers.  Diamond tweeter components were envisaged and modelled as early as 1987 (Fujimori 1998).  The ideal tweeter acts as a perfect piston, without deformation, at all frequencies.  The upper limit to the frequency at which the dome behaves ideally is called the break-up frequency.

With standard materials, modified geometries, such as a 20% reduction in diameter, can significantly increase the break-up frequency.  However these smaller units will be unable to achieve the sound pressure levels needed in high end audio systems.  For a given geometry, the break up frequency depends on the material (table 6) and is proportional to the sound propogation velocity. This velocity is proportional to $\sqrt{\dfrac{E}{\rho}}$, where E is Young's modulus and $\rho$ the density; diamond is expected to offer performance exceeding that of current materials by a factor of three.

**Table 6.** Properties of materials used in acoustic applications.

| Material | E(GPa) | $\rho$(kg/m$^3$) | $\sqrt{(E/\rho)}$ | Relative |
|----------|--------|------------------|-------------------|----------|
| Aluminium | 71 | 2700 | 5128 | 1.0 |
| Titanium | 120 | 4500 | 5164 | 1.0 |
| Beryllium | 318 | 1850 | 13111 | 2.6 |
| Diamond | 1000 | 3500 | 16903 | 3.3 |

In practice, a break-up frequency over 70 kHz has been achieved for diamond tweeters (figure 5), compared to 30 kHz for high quality aluminium tweeters. Cone breakup modes are always associated with a significant increase in harmonic and non-linear distortion, also with undesired lobing of sound pressure. In terms of sonic quality therefore cone breakup modes should be pushed as far away as possible from the audible range to avoid any kind of subharmonic interference, and hence the requirement on a high break-up frequency. For diamond to be viable for consumer acoustic applications, a key issue has been the ability to grow non-planar, near net shape, very thin components in high volumes at a low cost. Diamond tweeters are now being used in significant volumes (www.bowers-wilkins.co.uk); when assembled as part of B&W loudspeakers, they have met with expert and consumer approval (Messenger 2005).

## 4. Thermal properties and applications of diamond

The thermal conductivity of diamond is higher than any other material at room temperature, and contributes significantly to its performance in a number of applications including windows for multi-kilowatt $CO_2$ lasers (Godfried *et al.* 2000), megawatt gyrotrons (Thumm 2007) and even cutting tools. Its high thermal conductivity is a factor in predicting favourable figures of merit for active electronic devices (table 10) and optical components (table 8). A thermal conductivity of 2000 W m$^{-1}$ K$^{-1}$ at room temperature exceeds that of copper by a factor of five. Even lower grades with thermal conductivities of 1000 W m$^{-1}$ K$^{-1}$ exceed other competitor ceramics such as aluminium nitride by a factor of four to six.

The main heat conduction mechanism for many materials is by electrons; high thermal conductivity is associated with high electrical conductivity. In diamond, lattice phonons provide the main mechanism. Therefore any lattice defect acts to reduce its thermal conductivity by scattering phonons. Studies on isotopically controlled HPHT diamond grown by General Electric showed that the presence of $^{13}$C at its natural abundance (1.1%) reduced the thermal conductivity by about 30%.

As an element used purely for thermal management, natural diamond was used in some early microwave and laser diode devices (Seal 1971, Doting and Molenaar 1988). However, the availability, size and cost of suitable natural diamond plates limited its penetration into that market. With the advent of polycrystalline CVD diamond with thermal properties similar to those of type IIa natural diamond (figure 6) the availability problem was solved. Today, a range of thermal grades of diamond are available off the shelf from a number of sources. Since polycrystalline diamond is grown in large wafers, size is no longer limited to single devices or small arrays, but can extend to arrays many centimetres across. Consequently, CVD diamond has proven practical, being built into many devices from the 1990s onwards.

**Table 7.** Properties of selected materials used in laser diodes (room temperature values). The ranges for Diamond and AlN cover those available commercially. Data sources include Hudgins (2003)

| | Thermal conductivity (W m$^{-1}$ K$^{-1}$) | Coefficient of thermal expansion (ppm K$^{-1}$) | Young's modulus (GPa) |
|---|---|---|---|
| Diamond (polycrystalline) | 1000-2200 | 0.9 | 1100 |
| AlN (polycrystalline) | 170-285 | 4.2-5.3 | 308 |
| GaAs | 55 | 5.7 | 86 |
| GaN | 130 | 3.2-5.6 | 180 |

While thermal pyrolitic graphite (TPG) offers in plane conductivities in the range 1350-1700 W m$^{-1}$ K$^{-1}$, its low through-plane conductivity (around 98% less than the in-plane value) reduces its ability to get heat away from the active area of devices. In comparison high quality polycrystalline CVD diamond has been shown to have a through plane conductivity varying by less than 10-20% of its in-plane value (Twitchen *et al.* 2001).

### 4.1 Thermal integration considerations

For successful integration of a thermal management element into a device, the complete thermal path must be considered, along with the electrical requirements and thermally generated stresses. While diamond's extreme stiffness and low coefficient of thermal expansion are excellent for its use in high power transmission windows, they tend to count against it in purely thermal applications as they are significantly different to those of commonly used compound semiconductors (table 7). Thermally induced stresses can lead to reductions in device lifetime and reliability unless they are managed by, for example, pre-cracking the compound semiconductor (Heinemann *et al.* 1998) or using a diamond sandwich, where the upper layer acts to balance stresses.

Most active electronics packages require an electrically insulating layer between any metallisation and active cooling channels. Polycrystalline diamond is able to provide this with measured breakdown fields of 30-40 V/µm in commercial thermal grades (Element Six unpublished), and over 100 V/µm in research grades (Beuille *et al.* 2002).

When integrating diamond into a device package, the ideal geometry depends on factors ranging from the power density through to the location of cooling channels. However, this is readily modelled. For example, adding CVD diamond to the top of a microchannel cooled copper block reduces the device temperature rise from 22°C to 16°C as shown in figure 7. Varying the diamond geometry on a simply cooled copper block for a laser diode array (peak power densities of 100 Wmm$^{-2}$ at 200 µm spacing), suggested that the majority of the benefits occur with a diamond thickness of around 300 µm and 3 mm width. It should be noted that comparison of modelling with experimental results suggests that the metallisation is also an important component in the thermal path. A typical metallisation would be Ti/Pt/Au, with a total thickness of 1000 nm (Hall 2009). As well as providing physical attachment and one electrical connection, this provides intimate contact between heat source and heat spreader.

### 4.2 Integrated thermal management for RF devices

It has long been recognized that the diamond layer should be as close as possible to the active region of a high power device for the best performance. For example Touzelbaev and Goodson (1998) reported diamond deposition directly on a GaAs device. This approach has continued to concentrate on low temperature deposition, primarily of nanodiamond (Xiao *et al.* 2004).

Devices made in GaN have emerged as a technology of choice for next generation high power RF devices. Research exploiting diamond layers for thermal management has developed into two main areas: epitaxy-ready (111) composite silicon wafers with buried diamond layers (Zimmer 2007) and direct bonding of GaN to diamond wafers (Felbinger *et al.* 2007). A third possibility of growing RF device grade GaN directly onto a diamond substrate has not yet been successful (Hageman *et al.* 2003).

Integration of diamond with silicon enables access to standard GaN growth techniques. High electron mobility transistor (HEMT) devices manufactured using Group 4 Labs' material (figure 7) have been reported to demonstrate a gate temperature rise half that of the equivalent device fabricated on SiC (Felbinger *et al.* 2007).

## 5. Optical Properties

Diamond exhibits a wide range of desirable optical properties which, when combined with its excellent thermal properties, have led to a number of optical applications. Typically, but not exclusively, these involve high power laser technology such as CVD diamond beam exit windows in $CO_2$ lasers (Godfried *et al.* 2000 and Massart 1996).

The optical properties of diamond have been comprehensively reviewed by Zaitsev (2001). Here we consider those properties which are of particular practical interest, and review several emerging technologies.

### 5.1 Absorption and scatter

The absorption coefficient for intrinsic diamond from the UV to long-wavelength IR is illustrated in figure 8 (reproduced from Collins (2001)). Wide spectral transparency can be seen, even extending to 500 µm (Dore *et al.* 1998) except for regions of absorption in the infra-red between around 2.5 to 6.5 µm (4000 to 1500 $cm^{-1}$), and below the bandgap at around 226 nm (Dean 1965).

Other infra-red transmitting materials, such as ZnS and ZnSe owe their transparency to heavy atoms and weak bonds, resulting in low phonon energies and thus a cut-off frequency in the far infra-red. Diamond with its low mass and rigid lattice possesses the highest fundamental phonon (absorption) frequency of any material, with the single phonon mode centred at 7.5 µm (1332.5 $cm^{-1}$). While the lattice remains defect free, and thus symmetric, absorption at this frequency is forbidden. The absorption in the infra-red is due to two- and three-phonon processes. These are temperature dependent (Mollart *et al.* 2001), and the absorption is successfully described by the model developed by Piccirillo *et al.* (2002).

Recent measurements of absorption by calorimetry at 1064 nm (Turri *et al.* 2007) yield values of 0.003 to 0.07 $cm^{-1}$ for single crystal CVD diamond, while the authors have found the absorption coefficient at 10.6 µm to be between 0.02 and 0.05 $cm^{-1}$ on both single and polycrystalline samples (Whitehead 2003).

Polycrystalline diamond causes significant Rayleigh scattering losses at wavelengths below approximately 1 µm (Sussmann 2000). Therefore single crystal diamond is generally required for applications in the UV to near IR. In high crystalline quality single crystal diamond, residual scattering losses are primarily due to surface roughness, ranging from 0.04 to 0.6% for 1064 nm photons (Turri *et al.* 2007).

Raman scattering in diamond results in a shift in energy equivalent to 1332 cm$^{-1}$, with the Raman line width (a common metric of crystalline quality) less than 1.8 cm$^{-1}$ in high quality samples (Zaitsev 2001).

## 5.2 Dielectric properties of diamond

For light between the band gap up to 25 µm, the refractive index of diamond can largely be described by the Sellmeier equation. Diamond has one of the highest refractive indices in the UV part of the spectrum with n = 2.60 at 266 nm, decreasing to n = 2.46 at 405 nm (Zaitsev 2001).

At microwave frequencies, as well as the refractive index, the dielectric loss factor (tan δ) is important for high power transmission applications, e.g. as the output window of a gyrotron. These parameters affect reflection and power absorption. Sussman (2000) discusses the specific case for gyrotron tubes operating in the 70-170 GHz frequency range with output powers in excess of 1 MW for fusion research. At 145 GHz, dielectric loss tangent values as low as (0.6 ± 0.5) ×10$^{-5}$ have been reported on a 100 mm diameter polycrystalline CVD window (Heidinger *et al.* 1998). While this can be higher than for gold doped silicon and sapphire, diamond's dielectric properties are relatively insensitive to temperature change. The combination of its properties offers the highest performance for high power microwave transmission at both room and liquid nitrogen temperatures.

## 5.3 Luminescence

The luminescence spectrum of diamond is significantly altered by the incorporation of impurities into the diamond lattice. Each type of defect can give rise to spectral features from the UV to the IR due to electronic and vibronic transitions. Indeed the huge range of defects means that optical spectroscopy of diamond is a very rich subject area (Wilks and Wilks 1991, Clark *et al.* 1992, Zaitsev, 2001).

Biological applications which involve fluorescence imaging are particularly sensitive to the luminescence of the CVD diamond (Rittweger *et al.* 2009) Two basic techniques can be used to produce samples with low luminescence:

(i)     Growth of diamond with very low defect uptake such as that required in electronic applications (section 7)
(ii)    Using growth or post-growth damage techniques to create a very high concentration of defects which, through absorption, can minimize the diamond excitation volume, as well as producing non-radiative decay paths to quench the defect luminescence

For optical applications of CVD diamond the most important defect centres are arguably those due to single substitutional nitrogen (N$_s$), nitrogen-vacancy (NV) complexes and dislocations. CVD diamond samples containing nitrogen typically exhibit a broad spectrum of orange-red luminescence under excitation. This NV-related luminescence is absent in very high purity CVD diamond in which the residual blue luminescence, centred around 440 nm, is mainly associated with dislocations.

## 5.4 Birefringence and optical strain

There is growing demand to use CVD diamond where, in addition to its low absorption and high thermal conductivity, low birefringence is also required

Due to its cubic lattice symmetry diamond is nominally an optically isotropic material: the refractive index should be independent of the polarization of light. However, within the strain field of a dislocation, or group of dislocations, the lattice is distorted. As such diamond typically exhibits a complex birefringence pattern determined by the density and spatial distribution of dislocations and other extended defects. For optical applications which are polarization-sensitive, any birefringence in an optical element leads to depolarization of the beam and subsequent losses. In order to use diamond in these applications, low birefringence single crystal CVD diamond has been developed (Friel *et al.* 2009).

Gaukroger *et al.* (2008) discuss the relation between dislocation structures and CVD diamond growth, and some methods by which their density can be reduced. Figure 9 gives an example of the birefringence of two CVD samples with relatively high and low dislocation densities. It can be seen that the birefringence is reduced from, in this case, $\leq 1.2 \times 10^{-5}$ to $\leq 3 \times 10^{-6}$ by reducing the dislocation density in the material. Through the use of carefully selected and processed substrates, and the use of optimized growth conditions, birefringence $\leq 5 \times 10^{-7}$ has been reported (Friel *et al.* 2009).

It should be noted that polycrystalline diamond cannot be used for low birefringence applications due to the stresses between grains.

*5.5 Emerging optical applications based on single crystal CVD diamond*

The development of low birefringence CVD diamond has led to improvements in the performance of doped dielectric and semiconductor disk lasers. It has been shown that diamond heat spreaders bonded to the pumped surface of doped dielectric gain crystals allow pump power densities to be employed, and hence output powers achieved, that would otherwise cause the laser crystals to fracture (Millar *et al.* 2009). Intra-cavity diamond heat spreaders have also proven to be vital in realising the spectral coverage promised by semiconductor disk lasers at watt-level powers (Kemp *et al.* 2008). In particular, Hopkins *et al.* (2008) used this approach to improve output power by more than two orders of magnitude in the technologically important 2 - 2.5 μm band.

In contrast to either natural or high birefringence synthetic diamond, Millar *et al.* (2008) demonstrated that low birefringence diamond led to much lower depolarization losses. Single crystal CVD diamond is therefore a viable option for thermal management in laser systems of this kind.

Another area of recent development enabled by the availability of good optical quality, large area CVD diamond, is the use of diamond in Raman lasers as reported by Mildren *et al.* (2008). Here diamond is used as the active laser medium – the incident pump light is Raman-scattered to a new wavelength, whereby this scattered light is amplified (typically within a cavity) by the third-order nonlinear optical process of stimulated Raman scattering.

**Table 8:** Characteristics of selected Raman crystals. Kemp's figure of merit indicating resistance to thermal lensing (Kemp *et al.*, 2009) calculated using 8 mm low birefringence path length for diamond (Friel *et al.* 2009) and 25 mm for other materials.

| Material | Diamond | KGW | YVO$_4$ | Ba(NO$_3$)$_2$ |
|---|---|---|---|---|
| Raman Gain (cm/GW) | ~15 | ~4 | 5 | 11 |
| Raman Shift (cm$^{-1}$) | 1332 | 901 | 892 | 1047 |
| Thermal Conductivity (W/m.K) | 2000 | ~3 | 5.2 | 1.2 |
| Thermal Lensing Figure of Merit (for 1 μm light) | 1440 | 3 | 20 | 1 |

The attraction of diamond as a Raman gain crystal can be seen in table 8, despite the comparatively limited sizes of diamond available. The figure of merit for diamond is around two orders of magnitude higher than for conventional Raman crystals, largely due to its thermal conductivity. It also offers the largest wavelength shift.

The current record output power for a CW intracavity Raman laser (3W using KGW) is limited by the onset of excessive thermal lensing. The use of diamond, with its higher resistance to thermal lensing, may permit scaling into the tens of watts range.

### 5.6 Diamond Processing
In addition to the improvements in material quality for optical applications, developments in the processing and etching of diamond have opened up the possibility of fabricating structures in diamond such as micro-lenses (figure 10), gratings, waveguides or cavity resonators. Unlike other processes such as focussed ion-beam milling, etching using inductively-coupled plasmas is thought not to damage the diamond surface and sub-surface regions. Unlike typical oxygen-based plasmas, the argon-chlorine based plasma (Lee *et al.* 2008) has also been shown not to preferentially etch defects such as dislocations, and thus avoids the formation of undesirable etch pits.

## 6. Electronic properties and applications
Diamond is a wide bandgap semiconductor with an indirect gap of about 5.47 eV. One of the benefits of this is that diamond can support high electric fields before breakdown. Experiments on high purity CVD diamond have reported high mobility values and long lifetimes for electrons and holes. Combined with the high thermal conductivity, diamond can be the preferred material for a number of demanding electronic applications.

### 6.1 Carrier properties
6.1.1 *Intrinsic diamond.* The concentration of electrons in the conduction band of an intrinsic semiconductor, assuming a mid-bandgap Fermi level, is approximately $n = N_c \exp[E_g/(2k_BT)]$, where $N_c$ is the effective density of states in the conduction band, $E_g$ is the energy gap, $k_B$ is Boltzmann's constant and T is temperature. At room temperature $N_c \sim 2 \times 10^{19}$ cm$^{-3}$ (Nebel and Stutzmann 2000) and n is very much less than one electron per km$^3$. As perfect diamond has not yet been grown, the electronic properties will always be governed by impurities such as nitrogen and boron, and structural defects including dislocations and grain boundaries.

High carrier mobility in high purity intrinsic CVD diamond was first reported in 2002 by Isberg *et al.*. Recently, a number of other research groups have reported intrinsic CVD diamond with unintentional defect density around or below the parts per billion level ($\sim 10^{14}$ cm$^{-3}$) (Secroun *et al.* 2007). In the high carrier injection regime (Q>CV, where Q is the charge, C is the capacitance and V is the voltage), the mobility decreases with a $T^{-3/2}$ dependence consistent with acoustic phonon scattering up to temperatures $\sim$400 K. At higher temperatures, it shows a $\sim T^{-3.7}$ dependence consistent with measurements on natural diamond. At room temperature a hole mobility of 3800($\pm$400) cm$^2$V$^{-1}$s$^{-1}$ was observed. Even at 400 K the mobility still exceeds 2000 cm$^2$V$^{-1}$s$^{-1}$ (Isberg *et al.* 2002).

The combined electron and hole room temperature mobility for a number of electronic materials is plotted against bandgap in figure 11. Diamond's unique position is apparent, especially when thermal conductivity (indicated by the area of the circle) is considered.

Diamond's low relative permittivity and large bandgap combine to predict a widely quoted breakdown field of 1000 V/µm. Breakdown voltage measurements have been performed for Schottky diodes. For lateral diode structures, a breakdown voltage of 6 kV has been attained using a wide contact spacing (Butler *et al.* 2003) and up to 66 V for micron-scale separation devices. This second result corresponds to a simply calculated 146 V/μm breakdown field (Teraji *et al.* 2007). In vertical device structures, 2.5 kV breakdown has been reported for 18 μm thickness of intrinsic diamond (140 V/μm) (Twitchen *et al.* 2004). Others have also achieved breakdown fields of this magnitude (Huang *et al.* 2005, Kumaresan *et al.* 2009). Twitchen *et al.* (2004) reported 400 V/μm for a pressure contact; the authors have recently attained 450 V/μm in a vertical device structure (unpublished).

*6.1.2 Doped diamond.* As a dopant in silicon, boron provides an acceptor level 0.045 eV above the valence band. Dopants in wide bandgap semiconductors tend to have higher ionisation energies, resulting in low activation at room temperature. For example, 4H-SiC has shallow donors (n-dopant), but lacks a really shallow acceptor (p-dopant), the shallowest being aluminium with an ionisation energy of 0.19 eV (Ikeda *et al.* 1980). In the case of diamond, the range of dopants is limited, partly by its tight lattice spacing, and all have even higher ionisation energies (table 9).

While boron and phosphorus have been considered for high temperature applications (Kohn *et al.* 2004) and in *pn* junction UV diodes (Koizumi *et al.* 2001), their levels are too deep for conventional room temperature applications. Nitrogen is even deeper, so research has focused on unipolar devices using boron. However for boron to be active at room temperature its concentration must be pushed to very high levels, typically $>10^{20}$ cm$^{-3}$. As the concentration increases the conduction mechanism changes, first from band type conduction to hopping, then to metallic-like with the activation energy approaching zero (Borst and Weiss 1996, Lagrange *et al.* 1998). At the same time, the presence of compensating donors (shallow or deep) must be minimized.

**Table 9.** Common doping impurities introduced into synthetic diamond.

|  | Element | Activation Energy (eV) |
|---|---|---|
| N-type | Nitrogen | 1.7 |
|  | Phosphorous | 0.6 (Koizumi *et al.* 1997) |
| P-type | Boron | 0.37 (Collins *et al.* 1965) |

Reports claiming that sulphur (S) or oxygen (O) constitute reasonably shallow donors remain to be substantiated at the time of writing. A boron-hydrogen complex has also recently been reported to constitute a reasonably shallow n-dopant (Chevallier *et al.* 2002), but the long-term stability of this complex is uncertain. Goss *et al.* (2004) have performed a theoretical analysis of potential diamond dopants.

Implantation techniques to introduce dopants into diamond have had limited success (Prins 2003, Kalish 2003). Diamond's metastable properties at room temperature and pressure mean that implantation creates damaged or graphitic regions. While post implantation annealing can remove the majority of this damage, it cannot restore the diamond to a perfect lattice.

*6.2 Electrical contacts to diamond*

For diamond to be used as a semiconductor in devices, high quality contacts are required. Most metal contacts to semiconductors are non-ohmic due to surface depletion. The most common method used to achieve ohmic behaviour is to increase the doping concentration, and this has been employed in diamond. This has the effect of reducing the depletion width and hence the contact resistance by increasing tunnelling through the Schottky barrier. It has been found that by using carbide-forming metals (such as Mo, Ta, Ti) in combination with annealing, it is possible to make ohmic contacts to p-type diamond (Hewett and Zeidler 1992, Tachibana *et al.* 1992) with specific contact resistance as low as $\sim 10^{-7}$ Ω cm$^2$ (Werner *et al.* 1996). This is thought to be due to enhanced conduction through areas in which metal-carbide formation has occurred. Ohmic contacts to n-type diamond have been much more difficult to realize. Kato *et al.* (2009) reported Ti contacts to heavily phosphorus-doped diamond films ([P] $\sim 10^{20}$ cm$^{-3}$) that, although non-ideal, exhibited contact resistances down to $10^{-3}$ $\Omega$ cm$^2$.

Schottky contacts to diamond have been fabricated using both non-carbide forming (e.g. Al, Pt) and carbide forming metals, although the latter metals should not be annealed. Metal contacts to H-terminated electronic surface devices such as surface field-effect transistors often use Al for the gate and Au for the source and drain (Ueda *et al.* 2006). A more in-depth review of electrical contacts to diamond, including a consideration of the importance of diamond surface termination on the Schottky barrier height, is given by Werner *et al.* (2003).

*6.3 Diamond electronics – applications and markets*
Si and GaAs devices dominate the solid-state microwave device market today, however they have not yet been able to replace vacuum tubes in many high power applications. These tubes, such as klystrons and travelling-wave tubes (TWTs), are usually bulky, fragile and expensive. Wide bandgap materials (SiC, GaN, Diamond) may have the potential to replace TWT technology, in certain applications.

A further potential use of diamond is to control power at high voltages. A single diamond switch can in theory be used to switch power at voltages approaching 50 kV. This is not achievable with any other electronic material. If realized, electronic devices made from diamond could drastically reduce the size and cost of electronic control nodes, needed for active power grids.

**Table 10.** Material properties and figures of merit (normalized to Si) at room temperature for Si, 4H-SiC, GaN and diamond. 4H is the polytype of SiC that is considered best suited for power electronic devices (highest mobility). The diamond values are those reported for research grade single crystal CVD diamond in the space charge limited regime (Isberg *et al.* 2002). JFM is a measure of the high frequency capability of the material and KFM provides a thermal limitation to the switching behavior of transistors. High figures of merit do not mean that it will be straightforward to achieve these in practice.

|  | Si | 4H-SiC | GaN | Diamond |
|---|---|---|---|---|
| Band-gap (eV) | 1.1 | 3.2 | 3.44 | 5.47 |
| Breakdown Field (MVcm$^{-1}$) | 0.3 | 3 | 5 | 10 |
| Electron Mobility (cm$^2$V$^{-1}$s$^{-1}$) | 1450 | 900 | 440 | 4500 |
| Hole Mobility (cm$^2$V$^{-1}$s$^{-1}$) | 480 | 120 | 200 | 3800 |
| Thermal Conductivity (Wcm$^{-1}$K$^{-1}$) | 1.5 | 5 | 1.3 | 24 |
| Johnson's Figure of Merit (JFM) | 1 | 410 | 280 | 8200 |
| Keyes' Figure of Merit (KFM) | 1 | 5.1 | 1.8 | 32 |

*6.4  High power high frequency devices*
Due to physical limitations, silicon and gallium arsenide devices cannot achieve power levels above a few hundred watts (depending on the frequency to be amplified). Diamond, in principle, allows for higher power output per unit gate length at microwave frequencies. This is because a larger bias voltage, and hence the voltage amplitude on the microwave signal, can be supported across the transistor channel region over which the current is modulated.

The high ionisation energy of boron doped diamond means conventional device designs cannot automatically be expected to work well for diamond. One of the advantages of diamond is that the drift velocity of charge carriers reaches saturation at relatively low fields. This property may best be exploited by separating carriers from ionized acceptors (6.4.1), or by taking advantage of the surface conduction seen with hydrogen terminated surfaces (6.4.2).

*6.4.1  Bulk device structures.*  The *p-i-p* MISFET (metal-insulator field effect transistor) (figure 12a) was pioneered by Kobe Steel (Koizumi *et al.* 2002).  A trench is etched through the B-doped conducting channel; subsequently the gate dielectric and metal are deposited.  The current that flows around the gate is space-charge limited, which is not ideal for a high power transistor. Furthermore, successful operation requires very precise etching of the gate trench with nanometre control as the source-drain current is highly sensitive to the effective gate length ($L_{eff}$).

A more attractive design is shown in figure 12b which separates the holes and ionized acceptors by incorporating a 'delta' doping layer (a technique pioneered in silicon and commonly used in III-V devices).  The delta layer is a very thin highly boron doped region which acts as a source of holes.   Carriers can diffuse from the delta layer, with an expected mobility of around $10 \text{ cm}^2 \text{ V}^{-1} \text{ s}^{-1}$, and conduct travelling in the intrinsic layer, where mobilities in excess of $1000 \text{ cm}^2 \text{ V}^{-1} \text{ s}^{-1}$ are possible.   An enhanced variant of the single-delta MESFET (metal semiconductor field effect transistor) is the double-delta MESFET (figure 12c), where the additional delta-layer is deposited primarily to improve the source and drain contacts.  Recently, the University of Ulm demonstrated a diamond MESFET, incorporating a single delta-layer (figure 12b), that could operate at frequencies above 1 GHz (El-Hajj *et al.* 2008).  Diamond Microwave Devices Ltd. are working to commercialize devices of this type.

*6.4.2  Surface Conduction FET.*  An alternative design is based on transfer doping on the H terminated diamond surface. Hydrogen surface termination causes a 2D hole channel to form in the diamond layer several nanometres below the surface. It is believed that atmospheric contamination, especially condensation of water vapour onto the hydrogenated surface, plays a key role in creating the conductive channel below the surface (Ristein *et al.* 2008). During FET operation, holes travel through this channel from the source contact to the drain contact near the surface. The gate voltage modulates the hole conduction, making FET operation possible. Impressive results have been reported by several groups with headline figures including an output power density of 2.1 Wmm$^{-1}$ cw at 1 GHz (Kasu *et al.* 2007) and $f_t = 45$ GHz (unitary gain cut-off frequency) and $f_{max} = 120$ GHz (max frequency for operation) (Ueda *et al.* 2006). Although these results are promising, the long term stability of these surface FETs remains an issue, particularly at elevated temperatures and in harsh environments.

*6.5 Summary*
To realize the significant potential of diamond in devices outperforming existing technology will require: (i) access in volume to high quality, high purity, single crystal material; (ii) improvements in controlled doping; and (iii) the ability to process and contact thin layers and structures.

Access to bulk single crystal diamond for electronic R&D activities has already been largely achieved (Lohstroh *et al.* 2007 , Secroun *et al.* 2007, Nesladek *et al.* 2008).  Diamond Detectors Ltd. are commercially exploiting Element Six material for detector applications (next section). Dopants in diamond remain deeper than would be ideal, however it is anticipated that this can be overcome using smart designs.  Substantial progress has been made in developing CVD synthesis to fabricate these designs, and the recent achievement of operating frequencies above 1 GHz is extremely encouraging.

# 7.  Passive electronics

Diamond's role in passive electronics is primarily as a radiation detector. It excels in niche applications where volume sensitivity, radiation hardness and/or temperature insensitivity are required. In contrast to many other solid state radiation detectors, which use reverse biased diodes, high purity intrinsic diamond is able to act as a solid state ionisation chamber at room temperature. In order to do this charge must be able to travel freely through the lattice. The key to this is purity and crystalline quality.

Diamond's radiation hardness arises from its high atomic displacement energy (42 eV/atom) and low atomic number. It can demonstrate high sensitivity to radiation in comparison to other solid state detectors despite its relatively high electron-hole pair creation energy (13.2 eV, c.f. 3.6 eV for Si) because other detectors may have been intentionally radiation damaged to improve their effective radiation hardness (i.e. the stability of signal with time) (Buttar *et al.* 2000). In comparison to gaseous ionisation chambers, diamond's atomic density is of order 1000 times higher and therefore offers significant potential for miniaturisation. Since it has a large bandgap (indirect, 5.47 eV) it can demonstrate low leakage current, due to a low intrinsic carrier fraction, at room or even elevated temperatures. 5.47 eV is the energy of a 226 nm photon, therefore devices made using intrinsic diamond can be insensitive to visible light.

*7.1 Development of detectors*
*7.1.1 High energy physics.* In the early 1990s, the potential of CVD diamond as a radiation detector was explored, and compared to that of selected natural IIa diamonds. Early polycrystalline films achieved sufficiently high resistivity to be tested at fields in the region of 1 V/µm, and achieved mean carrier collection distances (CCDs) between 1 and 45 micron (Pan *et al.* 1992, 1993, Plano *et al.* 1994). Significant progress was made starting later that decade with the CERN RD42 group, concentrating on detection of ionizing particles for the high flux environment of the Large Hadron Colider (LHC). Ten years' work resulted in strip and pixel devices being manufactured on polycrystalline material 20 mm by 60 mm with a CCD of 300 µm (Adam *et al.* 2003, Wallny 2007).

Other high energy physics experiments such as BaBar at Stanford Linear Accelerator Center, Joint European Torus (JET), National Ignition Facility (NIF) at LLNL and Gesellschaft fuer Schwerionenforschung (GSI) have worked to develop diamond detectors. BaBar was an early user of diamond as a beam condition monitor (BCM) (Bruinsma *et al.* 2006). The role of a BCM is to reliably provide rapid beam abort signals in the event of beam instability and therefore protect other detectors from receiving damaging doses of radiation. The desired properties of a BCM are fast response (<2 ns pulse FWHM) and radiation hardness, which can be fulfilled in pieces of polycrystalline diamond approximately $10 \times 10$ mm$^2$ (Pernegger *et al.* 2004). Since BaBar, diamond BCMs have been installed in the ATLAS and CMS experiments at the LHC (Chong *et al.* 2007).

JET and LLNL are both fusion experiments and have worked to develop neutron detectors, sometimes with stringent timing requirements (Lattanzi *et al.* 2009, Schmid *et al.* 2003). As these are measurements on fast neutrons (MeV energies), the measurement can be made directly in diamond due to $^{12}$C(n, $\alpha$)$^9$Be reactions (Angelone *et al.* 2006). Absorption of the resulting $\alpha$ particle induces a charge that can be detected once it has drifted to the electrodes.

If the detection of thermal neutrons (tens of meV) is desired, thin layers of $^6$Li (usually in the form of $^6$LiF), $^{10}$B (as free-standing boron or boron doped diamond) or $^{235}$U can be applied as converter layers (Almaviva *et al.* 2008, Lardon *et al.* 2006).

*7.1.2 Measures of quality.* Electronic grade polycrystalline diamond achieving a CCD of 250 μm will typically be 500 μm thick (Pernegger 2006). An alternative measure for radiation detectors is charge collection efficiency (CCE).

$$CCE = \frac{CCD}{thickness}$$

CCE for the highest quality polycrystalline detector grade samples is therefore approximately 50%. More recently, high purity single crystal CVD diamond has been tested in detector applications (table 11). For this, the CCE can approach 100% for samples up to 800 μm thick (Pernegger 2006). This can occur because of a lack of traps, which therefore opens up possibilities of high resolution particle energy spectroscopy.

**Table 11.** Summary of properties of electronic grades of CVD diamond as typically used in radiation detectors (Pernegger 2006).

|  | Maximum CCD | Typical CCE | Typical operating field | $[N_s^0]$ [a] | $[B]$ [b] |
|---|---|---|---|---|---|
| Polycrystalline (pCVDD) | 300 μm | 50% | 1 V/μm | <50 ppb | <0.5 ppb |
| Single crystal (SC CVDD) | 800 μm | >95% | >0.3 V/μm | <5 ppb | <0.5 ppb |

[a] Measured by EPR: mean value for pCVDD, representative of whole sample for SC CVDD (Newton 2008).
[b] Total concentration measured by SIMS. Value given is equal to the detection limit (Chew 2007).

*7.1.3 Spectroscopy.* GSI are primarily interested in detection and analysis of heavy ion fragments, and have worked to develop detectors demonstrating fast signals, with high energy resolution, suitable for extreme radiation environments (figure 13). Testing single crystal diamond as a spectrometer, the FWHM measured for 5.5 MeV α particles was measured to be 17 keV, compared to 14 keV using a Si *p-i-n* diode. This corresponds to an energy resolution of 0.3% (Pomorski *et al.* 2006, Berdermann *et al.* 2008). These authors also demonstrated a timing resolution of 28 ps (one standard deviation for the correlation between two detector responses) for both polycrystalline and single crystal diamond.

Neutron spectroscopy measurements for fast 14 MeV neutrons have also been demonstrated (Schmid *et al.* 2004). An initial device measured an energy width of 240 keV for the 8.4 MeV α particles produced, a FWHM of 2.9%. However most of this width was attributed to the energy spread of the incident neutrons.

*7.1.4 Further uses.* Diamond can also be used to detect ionising radiation, such as x-rays and ultraviolet light. Similarly to α particles, sub-bandgap UV light has a short penetration depth into diamond and can be efficiently detected using thin devices, possibly with a surface electrode configuration. X-ray detection is due to volume excitation. Examples include quadrant detectors for synchrotron beam position measurement (Morse *et al.* 2007, Bergonzo *et al.* 2006) and medical radiotherapy dosimeters (Whitehead *et al.* 2001, Cirrone *et al.* 2006, Tranchant *et al.* 2008). UV detection has been studied for LYRA/BOLD (BenMoussa A *et al.* 2008).

Diamond can also be used in a similar manner to another important class of detectors, scintillators. In some extreme fluence environments such as synchrotron beams, the heat load can readily exceed the capabilities of traditional materials. Diamond's high thermal conductivity and known photoluminescence defects enable its use as a fluorescence detector in these circumstances (Bergonzo *et al.* 2006, Nam and Rijn 1992). Separately, controlled levels of defects within diamond can be used to give simple methods of integrating the radiation dose over time through thermal luminescence dosimetry (Marczewska *et al.* 2006).

### 7.2 Effects of imperfections

Since passive electronic detectors rely on transport and collection of electrons and holes, the nature and types of defect present strongly affect how well they work. The device performance depends on defects in the bulk and near the surface and electrical contacts.

Assuming high quality diamond growth, the surface defects are dominated by damage remaining due to surface processing. For most devices, the as-grown surface is either too rough (polycrystalline diamond) or still has a low grade substrate attached (single crystal devices) and will therefore require mechanical processing. However, while flatness of the final surface can be important to enable, for example, high spatial resolution, the flattest surface is not necessarily the best quality with regard to sub-surface damage. Post-processing etching has been developed and shows promise for some applications.

Establishing high quality contacts to the diamond is also important. Repeated re-contacting of the same diamond device has shown that significant variations in performance can occur, for the same contact recipe. Evaporated gold can work well in many circumstances but, in the authors' experience, can suffer from low mechanical adhesion and thus poor longevity. Other successful schemes include layers with carbide formers such as Cr or Ti, and one type of DLC developed by Diamond Detectors Ltd. (Galbiati A *et al.* 2009).

For many polycrystalline diamond detectors, "priming" has been shown to improve the device performance. This can be with x-rays or electrons - any radiation that creates charge carriers throughout the bulk of the diamond and can therefore fill all active traps (Pernegger 2006). While the grain boundaries are the most obvious defect, and have been correlated with reducing the CCE (Adam *et al.* 2003), the precise nature of traps filled (or de-activated) by priming has not been determined.

### 7.2.1 Radiation hardness.
Once low defect diamond has been grown and carefully processed, new defects can be created by radiation damage during use. Studies of device performance versus proton and neutron dose have been performed, including by RD42 for CERN/LHC, demonstrating its suitability for that extreme environment. These data and other studies covering a range of irradiation energies are summarized in table 12.

**Table 12.** Measured radiation hardness of diamond as a function of energy (10 to $10^4$ MeV) and irradiation type (protons or neutrons). Type and quality of diamond used in each study is indicated.

| | | Wallny (2007) | de Boer *et al.* (2007) | Lohstroh *et al.* (2008, 2009) |
|---|---|---|---|---|
| Protons | 50% Dose [a] | $6 \times 10^{15}$ p cm$^{-2}$ | $4 \times 10^{14}$ p cm$^{-2}$ | $2 \times 10^{15}$ p cm$^{-2}$ |
| | *Energy* | *24 GeV* | *26 MeV* | *2.6 MeV* |
| | | High quality pCVDD | Low quality pCVDD | High quality SC CVDD |
| Neutron | 50% Dose [a] | - | $1.25 \times 10^{15}$ n cm$^{-2}$ | $<2 \times 10^{13}$ n cm$^{-2}$ (100% signal) |
| | | | | $>1 \times 10^{16}$ n cm$^{-2}$ (Slight signal) |
| | *Energy* | - | *20 MeV* | *0.7 MeV* |
| | | | Low quality pCVDD | High quality SC CVDD |

[a] Irradiation dose observed to give 50% reduction in signal

The energy dependence of non-ionising energy loss (NIEL) is calculated by de Boer and colleagues (2007) and compared to experiment. It predicts a decrease in NIEL by two orders of magnitude, and thus an increase in radiation hardness, as the proton energy is increased between 10 and $10^4$ MeV. For neutrons, calculations show significantly lower energy deposition due to the reduction of charged particle induced damage, with a smaller energy dependence (less than one order of magnitude over the range 10 to $10^4$ MeV).

When designing for radiation hardness with lower incident energies (<10 MeV) it is important to consider where the energy is deposited. For lower proton energies, all particles are stopped within the detector. For example, 2.6 MeV protons only have a range of 40 µm, with most damage occurring very close to that depth. Even though the damaged region is very small, it can stop the device functioning effectively (Lohstroh *et al.* 2008). For neutrons with energies in the range 1-20 MeV the onset of inelastic reactions, such as $^{12}C(n, \alpha)^9Be$, as the energy increases, needs to be considered (Angelone *et al.* 2006). These reactions can provide another path for radiation damage to occur.

To date, recovering the performance of devices following irradiation has been confined to priming, akin to that used with polycrystalline diamond. The authors are not aware of published data on the nature of the defects, or their response to annealing.

## 8. Electrochemical properties and applications of diamond

In the field of electrochemistry, highly boron doped diamond has been used in electroanalysis and bulk oxidation of dissolved species in solution. Its chemical inertness, resistance to fouling and large potential window in aqueous solution make it a unique electrode material.

### 8.1 Chemical inertness of diamond

Diamond is extremely resistant to oxidation and attack by acids, even at elevated temperatures. Unlike nearly all other electrode materials, diamond is also resistant to attack by hydrofluoric acid. Platinum electrodes have been used for generating oxidising species in HF solutions for the etching of silicon in the absence of nitric acid. However, the solution eventually becomes contaminated with trace amounts of platinum, which in turn can contaminate the silicon surface. Replacing the platinum electrode with a diamond electrode has removed this contamination problem and in some cases the diamond electrode has been used to measure the extent of any contamination in the HF solution (Ponnuswamy *et al.* 2001).

Whilst the bulk of the diamond is inert the surface of the diamond can be altered. Upon removal from a CVD reactor the surface of diamond is hydrogen terminated and therefore hydrophobic. Under anodic operation (operation at positive voltages) in aqueous solution, this diamond surface rapidly transforms from hydrophobic to hydrophilic, becoming oxygen terminated.

### 8.2 Resistance of diamond electrodes to fouling

Some organic chemicals such as phenol have the ability to react with metal electrodes, resulting in the electrode surface becoming inactive. This is termed as 'fouling'. Diamond is thought to be much less prone to fouling (Hagans *et al.* 2001), although other authors note that de-activation due to fouling of the diamond does occur, unless the diamond anode is periodically run at a high potential (E > 2.3 V versus the standard hydrogen electrode (SHE)). Changing the polarity of the electrodes periodically can also prevent build up of foreign material on the diamond surface (Haenni *et al.* 2004).

Should the diamond surface become fouled in an industrial application, an acid flush can be used to re-activate the surface without degrading the diamond electrodes. In electroanalysis, where the electrode is more easily removed from the system, the electrode surface can be re-activated by polishing. There is little risk of damaging the electrode surface whichever abrasive is used.

### 8.3 Large potential window of diamond electrodes

Whether an electrode is used for electroanalysis or bulk oxidation of contaminants in solution, the potential which can be applied before electrolysis of the electrolyte/solvent occurs is an important property. Since most electrochemistry is performed in aqueous solution this corresponds to the potentials at which water is electrolysed into hydrogen gas (at the negatively charged cathode) and oxygen gas (at the positively charged anode).

Many groups have reported diamond electrodes with a large potential window for water stability and featureless background currents within this window (Swain and Ramesham 1993, Swain 1994a,b, Alehashem *et al.* 1995, Awada *et al.* 1995, Martin *et al.* 1999, Argoitia *et al.* 1996, Martin *et al.* 1995). Figure 14 shows a typical cyclic voltammogram for high quality boron doped diamond, with the flat region extending from approximately -1.35 V to +2.3 V.

In the case of electroanalysis the wider window may allow new systems to be investigated or new species to be detected. The low background currents within the window enable greater sensitivity than for metals such as platinum. In the case of bulk oxidation driving the system at higher potentials allows reactions to occur at higher rates (Carey *et al.* 1995).

Diamond remains inert in all known organic solvents and has been reported to display a similar potential window to that of glassy carbon in many (Yoshimura *et al.* 2002).

If significant amounts of $sp^2$ carbon are present in the film i.e. it is of poor quality, the potential window is significantly reduced, becoming similar to that of glassy carbon or highly oriented pyrolitic graphite (HOPG). There are also reports of polycrystalline diamond electrodes with large amounts of $sp^2$ carbon in the grain boundaries eventually disintegrating under anodic polarization (large positive voltages) (Angus *et al.* 2004).

### 8.4 Diamond electrodes for electroanalysis

As discussed in the review by Swain (2004), important characteristics in electroanalytical applications are the linear dynamic range, sensitivity, lower limit of detection, response time, response precision and response stability.

Two problems with $sp^2$ carbon electrodes, such as glassy carbon, are the stabilization time and the short and long term signal stability. Typically glassy carbon takes an initial 15-60 minutes to stabilize whereas diamond is usually in the 1-5 minute range. At potentials used for detection of redox species the surface of glassy carbon gradually oxidizes leading to an unstable background signal. For diamond, after the initial transformation to an oxygen terminated surface, the microstructure is extremely stable, giving very stable, low background currents.

Diamond has also an advantage over most other electrode materials for sensing in aggressive or corrosive environments.

To reduce background currents further, the surface area of the electrode should be reduced. If a microelectrode with a radius of only a few microns is used, diffusion to the electrode surface is no longer planar but hemispherical. This results in higher mass transfer rates to the electrode surface and faster response times (Forster 1994). A microelectrode array can increase the signal whilst still maintaining the advantages of the microelectrode.

Colley *et al.* (2006) have characterized microelectrode arrays manufactured entirely from diamond (figure 15). Being an all diamond structure, rather than having different material for the insulator, enables truly co-planar microdisk electrodes. It can also be cleaned with abrasives without producing any change in its structure.

Many papers have shown a variable response from diamond electrodes when used as electrochemical sensors. However Wilson *et al.* (2006) emphasize that high quality electrical contacts to the boron doped diamond are required to obtain repeatable results.

*8.5 Diamond electrodes for advanced oxidation*

Several organisations, including CONDIAS and CSEM are selling diamond electrodes for electrochemical water treatment. The majority of these electrodes consist of a thin coating of boron doped diamond (typically a few microns) on an electrically conductive substrate such as refractory metals (e.g. tungsten, molybdenum or tantalum) or a conductive ceramic (e.g. doped silicon or electrically conductive silicon carbide).

Another approach is to use thick, freestanding, boron doped diamond discs (see figure 16). Unlike diamond coatings they do not suffer from degradation due to pin holes. Such discs have been used in a system produced by Advanced Oxidation Limited (AOL).

Figure 16 shows results from a landfill leachate treated with the AOL reactor. Both the ammonia and the chemical oxygen demand (COD) are oxidized and the solution quickly becomes colourless.

Exactly what oxidants are produced in aqueous solution when a large potential is applied to diamond electrodes is not well known. Even less is know about the mechanism by which these oxidants are formed. However, it is believed that diamond electrodes produce a mixture of oxidants such as OH· (hydroxyl radicals), ozone and hydrogen peroxide.

Advanced oxidation processes (AOPs) are those using OH· as a strong oxidant. With an electrochemical oxidation potential of 2.80 V, OH· is only weaker than fluorine (3.06 V). It is significantly stronger than other oxidants such as ozone (2.08 V) and the hypochlorite ion (1.49 V).

The second order rate constants for reaction of hydroxyl radicals with most organic compounds are of the order $10^8$ to $10^9$ l mole$^{-1}$ s$^{-1}$. This is of the same order as acid-base reactions, which are considered to be one of the fastest reactions in solution.

Advanced oxidation processes are normally only applied to low COD wastewaters because of the cost of producing hydroxyl radicals. However, it may not be necessary to oxidize the COD completely. For example, material that was previously resistant to biological degradation can be treated with an AOP and then re-processed by biological means (Tchobanoglous *et al.* 2004, Crittenden *et al.* 2005).

The efficiency of AOPs can be affected by the presence of suspended solids and the pH of the wastewater. Also, high concentrations of carbonate and bicarbonate ions in the solution being treated cause a significant reduction in the amount of organic material being destroyed, due to unfavourable rate constants (Tchobanoglous *et al.* 2004, Crittenden *et al.* 2005).

*8.6 The future of diamond electrodes.*

The properties of diamond as an electrode clearly show that it is a unique material. The stability of the diamond surface, its chemical inertness and the ability to apply large potentials before the electrolysis of water occurs are all factors in favour of selecting diamond as an electrode material. Most electrochemical synthesis and processing applications demand high volume capability and high reaction rates, in turn requiring large area electrodes operating in very aggressive chemical environments. Diamond is unparalleled in its ability to survive such environments without erosion, and at the same time reduce the process cost. Trends in legislation are continually reducing the permitted levels of harmful chemicals in effluent, and CVD diamond electrodes offer promising solutions to these new challenges.

Electroanalysis applications have different requirements to those of chemical synthesis and processing: the key requirements are generally sensitivity and stability. Typical industries which might use electroanalysis such as the medical, pharmaceutical or bioscience sectors are currently growth areas where a new material can quickly establish itself if a novel use is found. In fact, the biocompatibility of diamond has already led to a novel use, for in vitro measurements of neurotransmitter levels, electrical activity and also for neural stimulation (Halpern *et al.* 2006).

## 9. Summary and outlook

Diamond is a unique material with many properties that cannot be surpassed. Its growth by CVD enables these properties to be optimized, controlled and exploited in many applications.

Its strong lattice is most obviously important in wear applications, but also feeds into its stiffness as an acoustic material and its radiation hardness for detectors. The lattice enables the phonon transport that gives diamond its exceptional thermal conductivity, and its strength helps create robust optical and microwave windows.

Transparency over such a wide wavelength range is technically fascinating, and extremely useful. Diamond became useful at long wavelengths when large areas and millimetre thicknesses were attained. Controlling absorption, photoluminescence and birefringence in single crystal diamond, by reducing impurities and defects, now opens up applications with visible light.

Its carrier transport properties and wide bandgap can be exploited in electronics, with reduced defect concentrations enabling operation at higher voltages, power and frequencies than can be achieved with other semiconductors. Controlled addition of impurities is also essential to provide carriers and low resistance contacts in these applications.

The wide electrochemical potential window of diamond can be utilized if the highest levels of boron are added, bringing new electroanalysis techniques and environmentally beneficial waste treatment.

Its high thermal conductivity is a key enabler for other technologies as a heat spreader. It underpins its potential performance in high power electronics and in novel visible and near infra-red lasers, as well as its delivered performance in abrasive tools, multi-kW laser windows and megawatt microwave windows.

In the last ten years advances in commercial CVD diamond growth have enabled consistent, high quality grades of diamond with tailored properties. Diamond is being used in applications ranging from tweeter domes, to exit windows for lasers used in industry.

Moving forward, each new user will set their priorities, but areas for continued development include electronic doping, increasing the size of single crystals and cost reduction. With access available to high quality material, new users will address challenges of integration to create novel devices. They will develop techniques to control strain and bonding in lasers, and enhance metallization and packaging for electronics. The potential of diamond extends far beyond the properties traditionally associated with mechanical applications or gemstones, and CVD diamond is already proving that it brings innovative solutions to a range of challenging applications.

# Figures

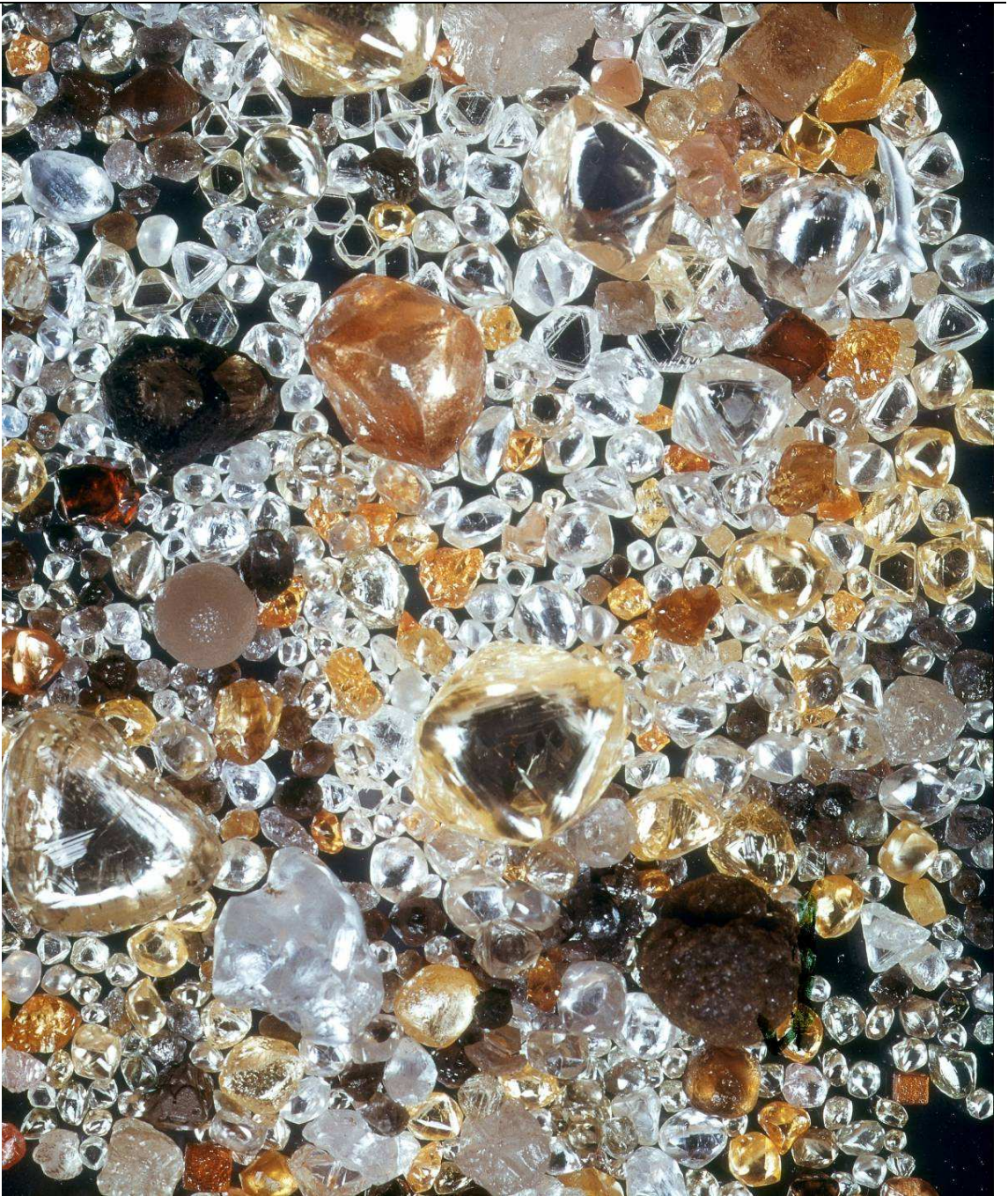

**Figure 1.** An array of natural diamonds showing a small cross-section of the infinite array of natural diamond's colours and shapes and consequently material properties.

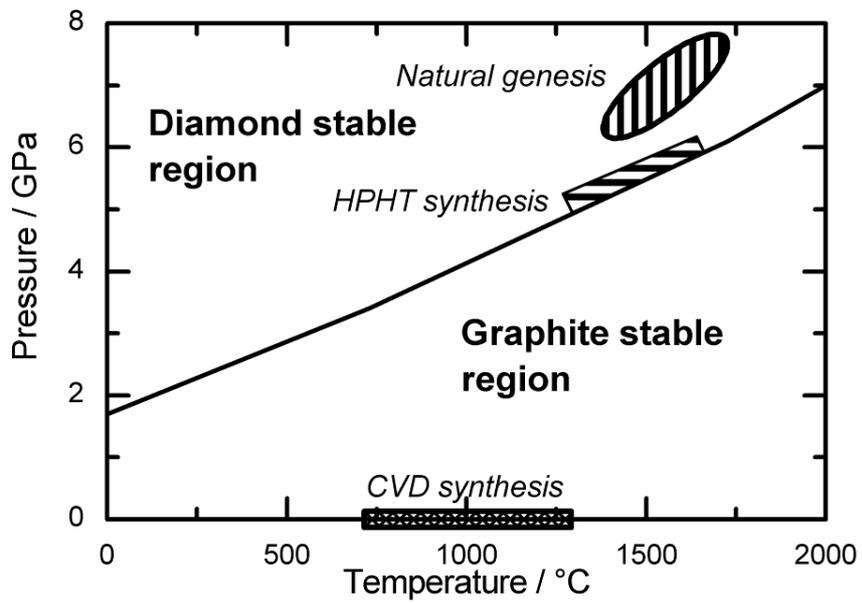

**Figure 2.** Phase diagram for carbon indicating main regions of pressure-temperature space in which diamond growth occurs.

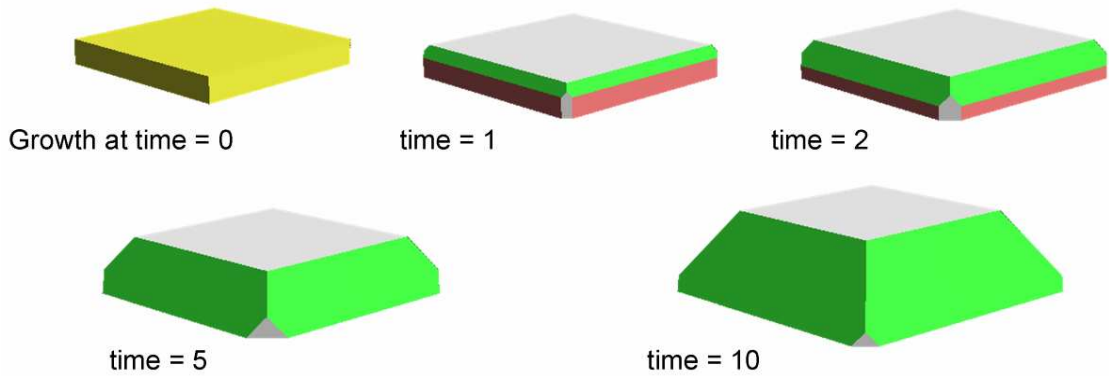

**Figure 3.** Theoretical morphology development under conditions of α = 4, with low β and γ from a {100} faced, <110> edged substrate (yellow). Non-{111} facets are eventually extinguished from the growth. Growth with {100} facet is shown in white, {111} in green and {110} in pink. (Morphology model courtesy of DTC Research Centre, Maidenhead).

**Polycrystalline, mechanical grade CVD**

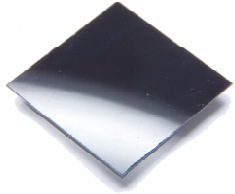

**Polycrystalline, optical grade CVD**

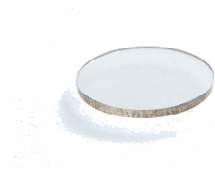

**Single crystal HPHT diamond**

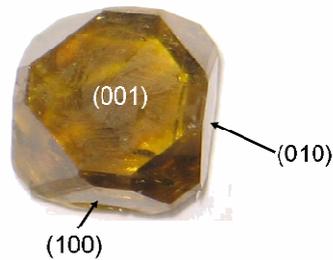

**Single crystal CVD diamond**

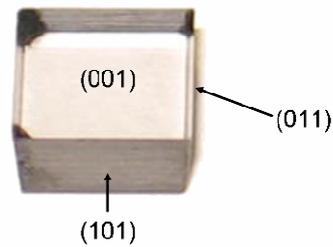

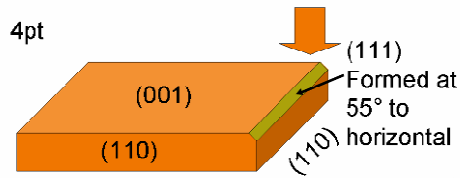

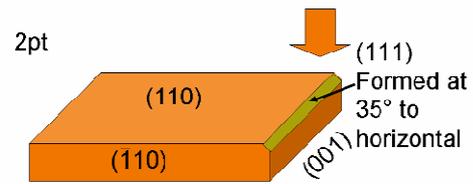

**Figure 4.** Mechanical (top left) and Optical (top right) grades of polycrystalline diamond. The laser cut edge of the optical sample highlights the grain structure of polycrystalline diamond. Middle: HPHT and CVD single crystal diamond used for tool applications.
Bottom: Typical geometries (4 pt – {100}, 2pt – {110}) used with synthetic diamond in tool applications. Arrows indicate the direction of the cutting edge.

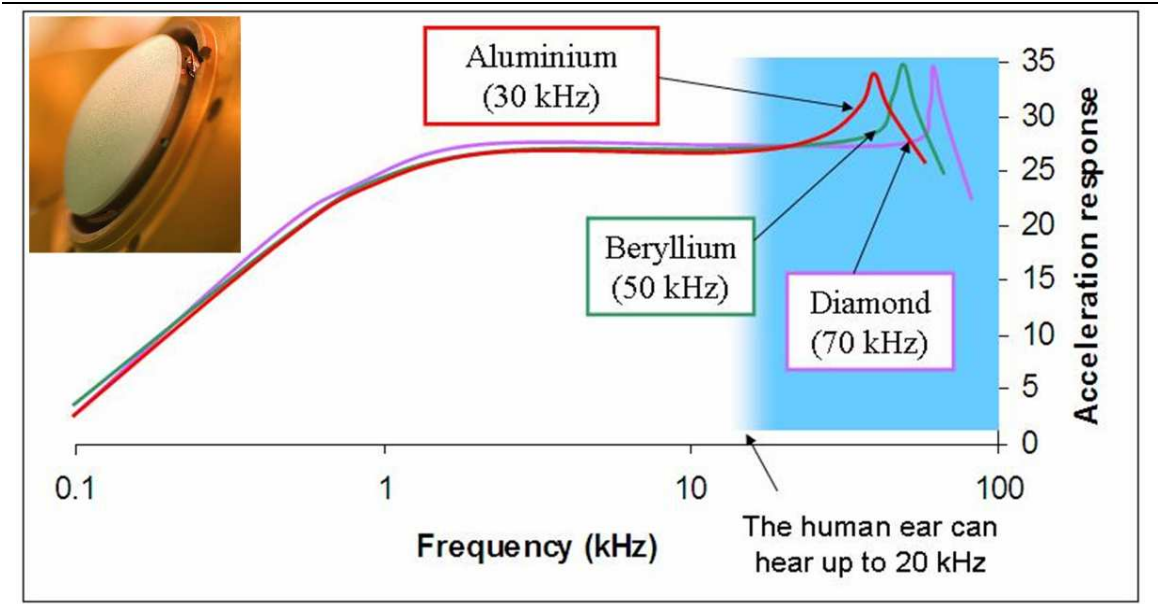

**Figure 5.** A 26 mm diameter polycrystalline diamond hi-fi tweeter speaker dome (inset) with a break up frequency of 70 kHz (c.f. 30 kHz for Aluminium and 50 kHz for Beryllium). Image and graph courtesy B&W loudspeakers.

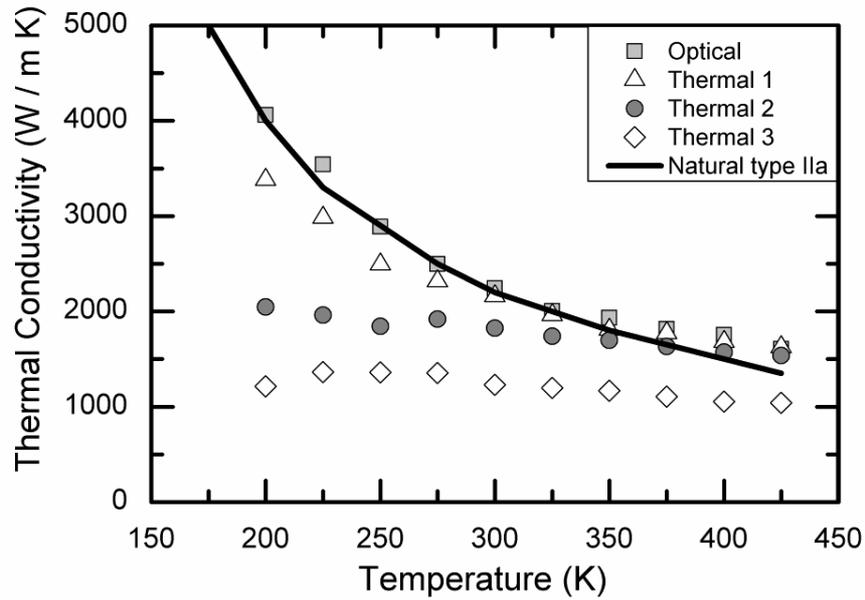

**Figure 6.** Thermal conductivity versus temperature for natural type IIa (Vandersande 1993) and CVD polycrystalline diamond. Thermal 1, 2 and 3 are representative of grades with nominal room temperature conductivities of 1800, 1500 and 1000 W m$^{-1}$ K$^{-1}$ respectively.

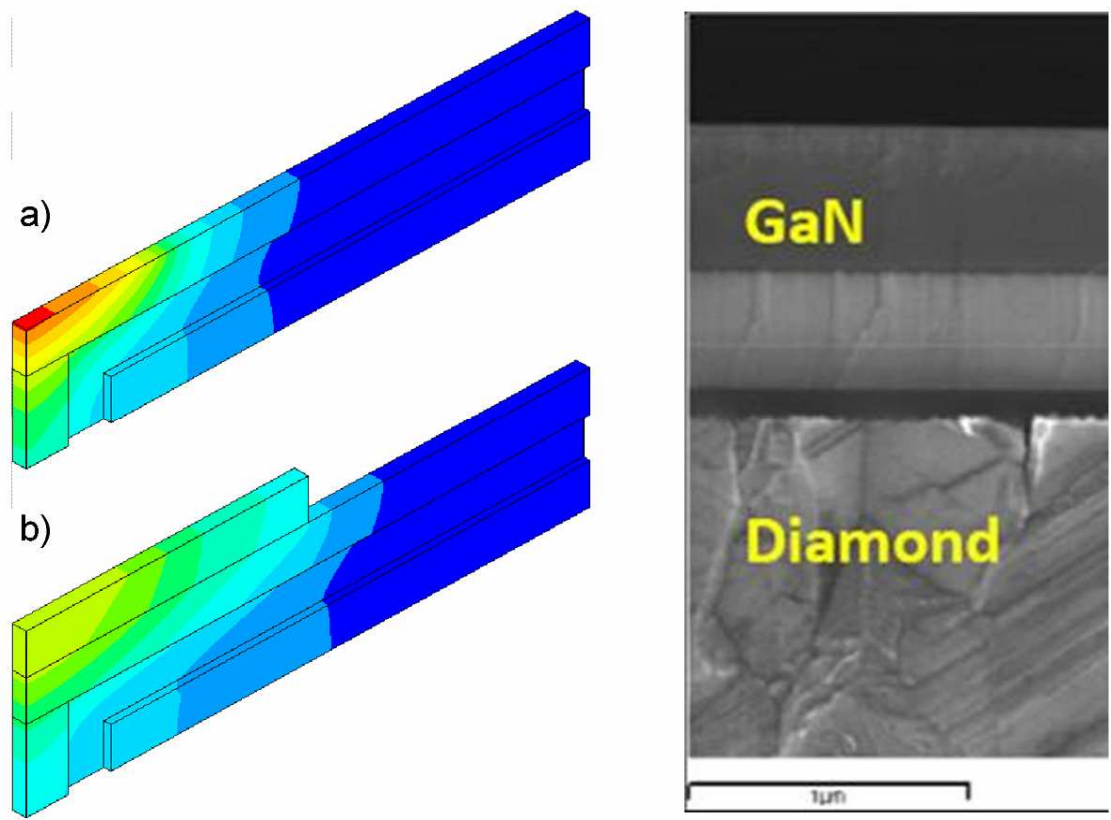

**Figure 7.** Left: model of copper microchannel heatsink (a) with no heatspreader, (b) with diamond heatspreader beneath the device. Peak temperature rise reduces from 22°C (red) to 16°C (green).
Right: Proprietary wafer bonding and release technology enables Group 4 Labs to attach 2 µm of device grade GaN to a CVD diamond substrate 100 µm in thickness.

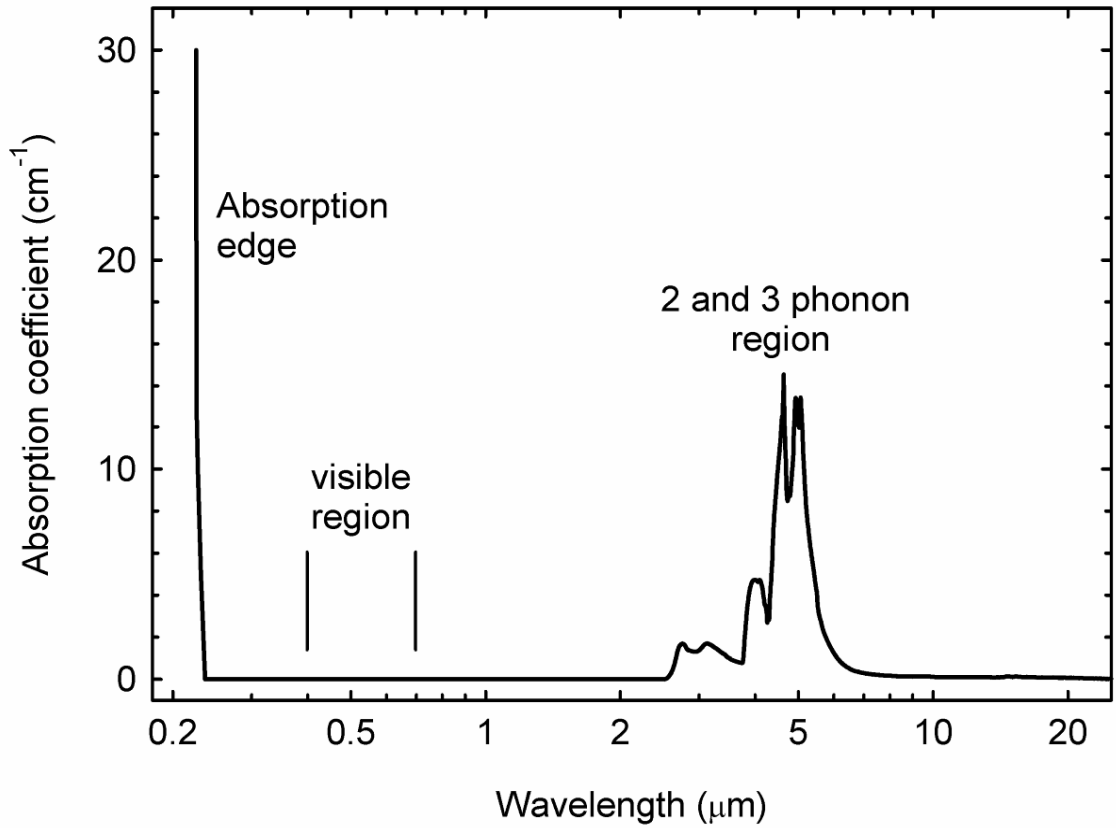

**Figure 8.** Absorption spectrum of high purity diamond from the UV to mid-IR (reproduced from Collins (2001))

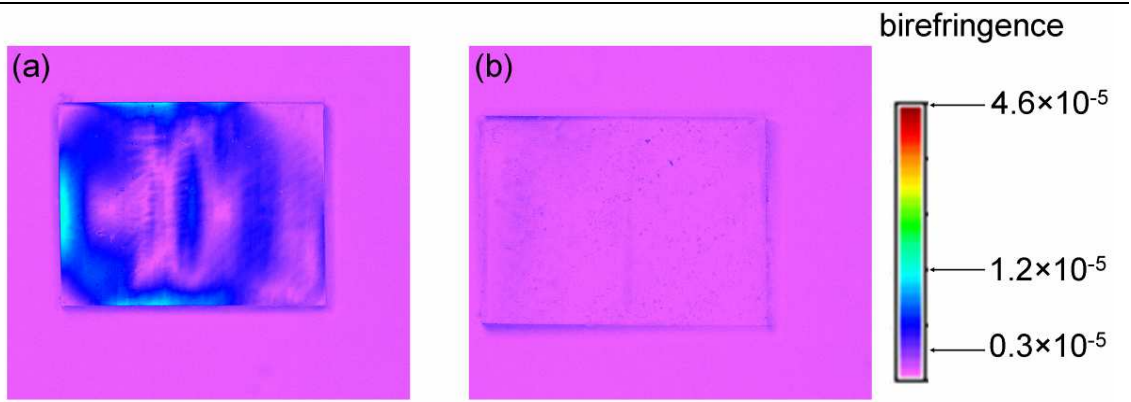

**Figure 9.** Birefringence micrographs for two CVD samples with (a) relatively high and (b) low dislocation density.

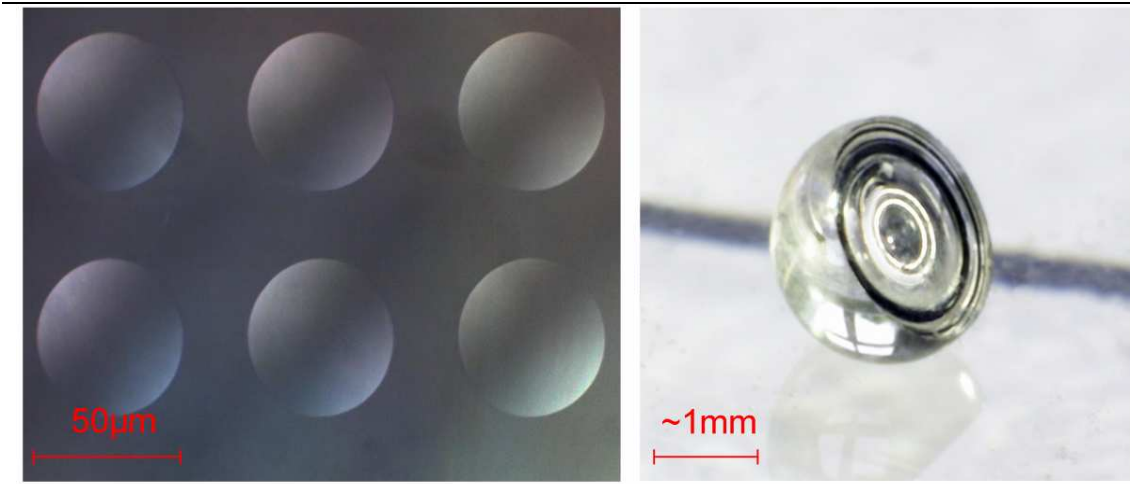

**Figure 10.** Left: Etched microlens array (Lee *et al.* 2008). Right: Element Six single crystal CVD diamond solid immersion lens.

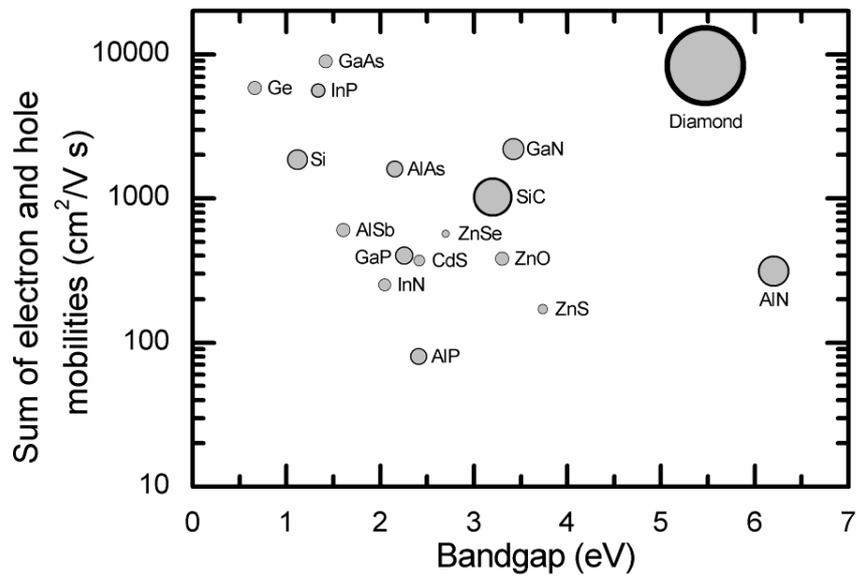

**Figure 11.** Combined electron and hole mobility (room temperature) for electronic materials. Diamond mobility from Isberg *et al.* (2002). Circles have an area proportional to the thermal conductivity.

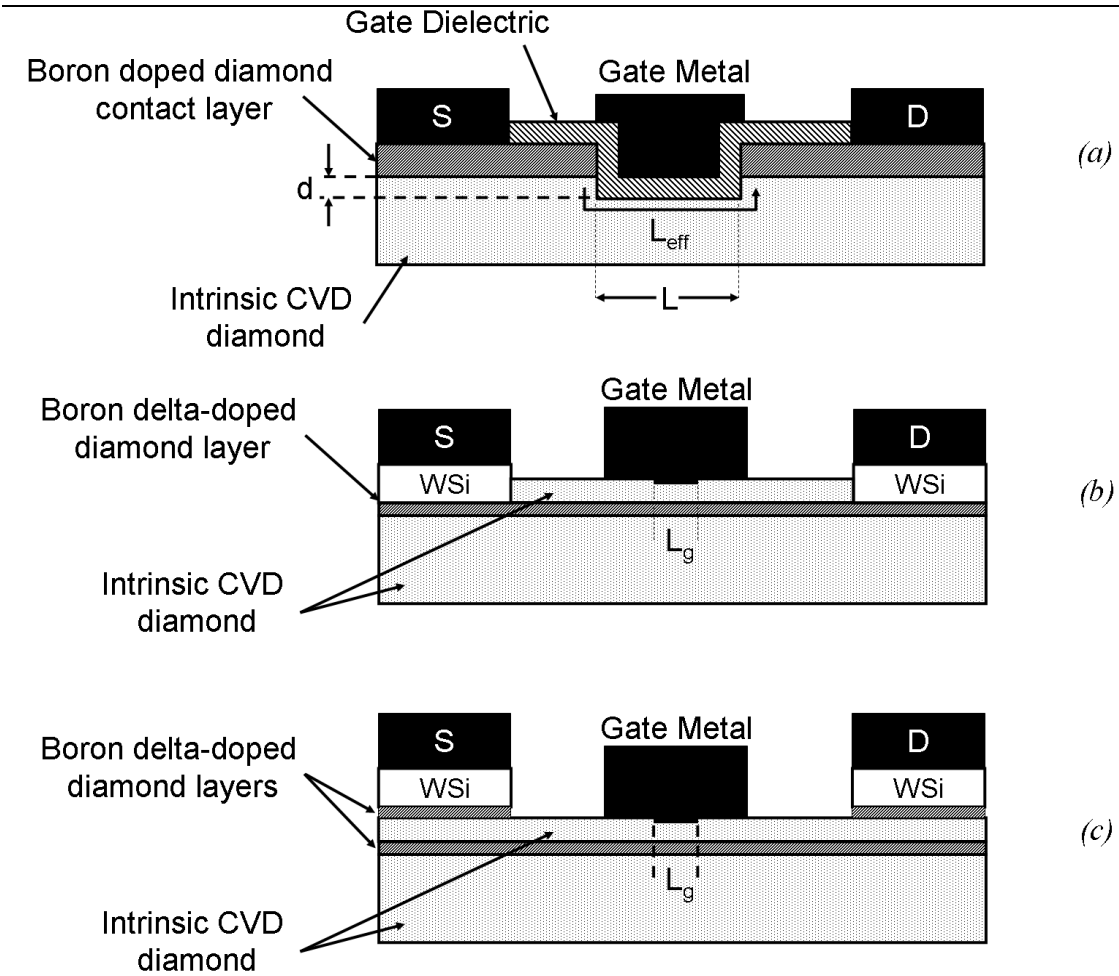

**Figure 12.** Diamond FET design concepts: (a) p-i-p MISFET (b) delta MESFET (c) double-delta MESFET. [S – source contact , D – drain contact, WSi – Tungsten-silicon contacting layers]

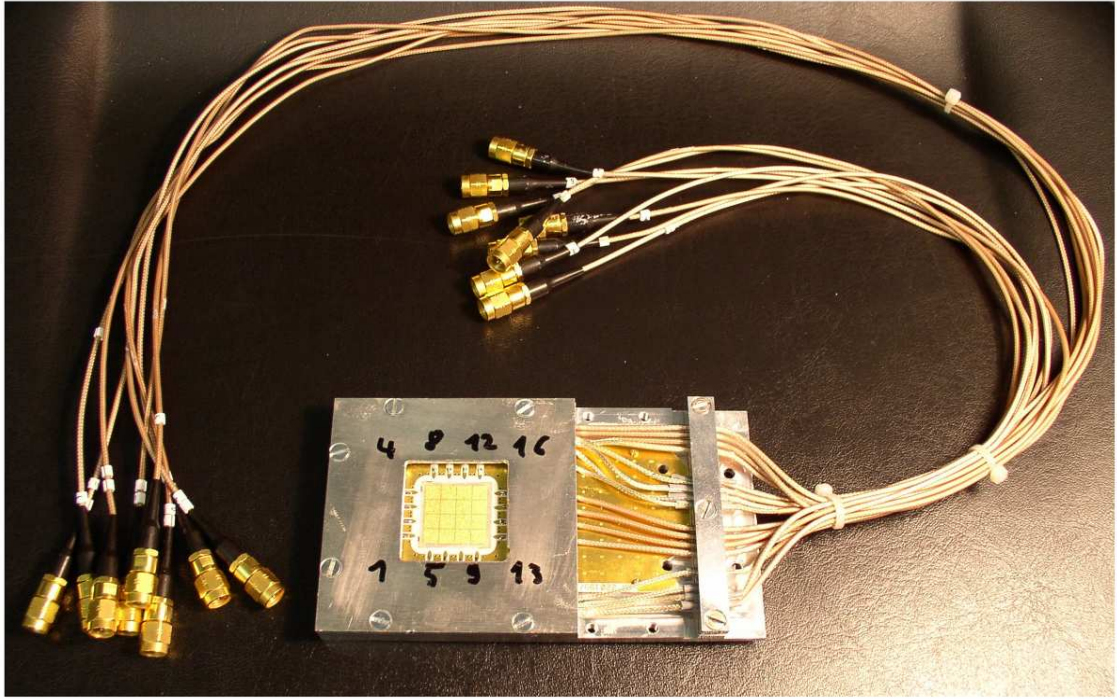

**Figure 13.** A macroscopic 16 pixel detector manufactured in polycrystalline diamond (image courtesy E Berdermann, GSI).

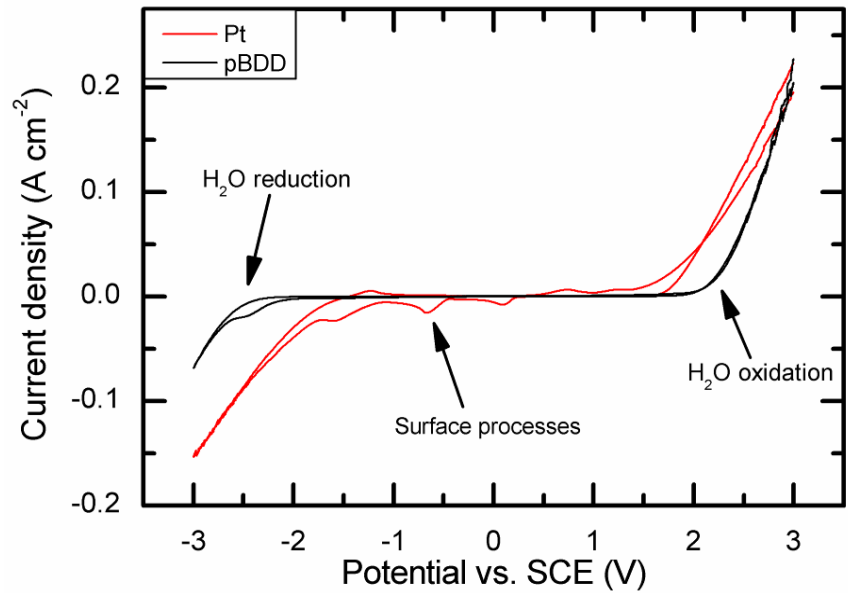

**Figure 14.** Cyclic voltammogram of high quality boron doped diamond (BDD) demonstrating wider potential window and lower background currents than platinum (SCE – saturated calomel electrode)

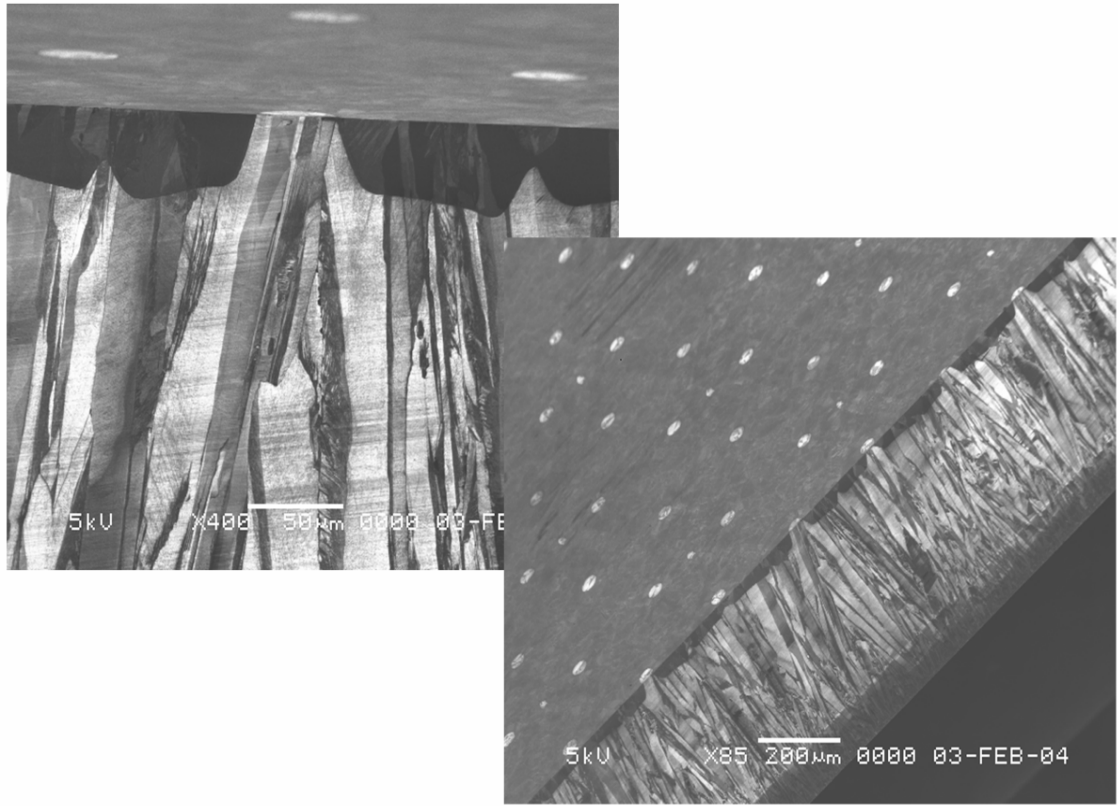

**Figure 15.** SEM images of an all diamond microelectrode array produced by Element Six. Boron doped diamond appears lighter. Upper figure reproduced from Pagels *et al.* (2005).

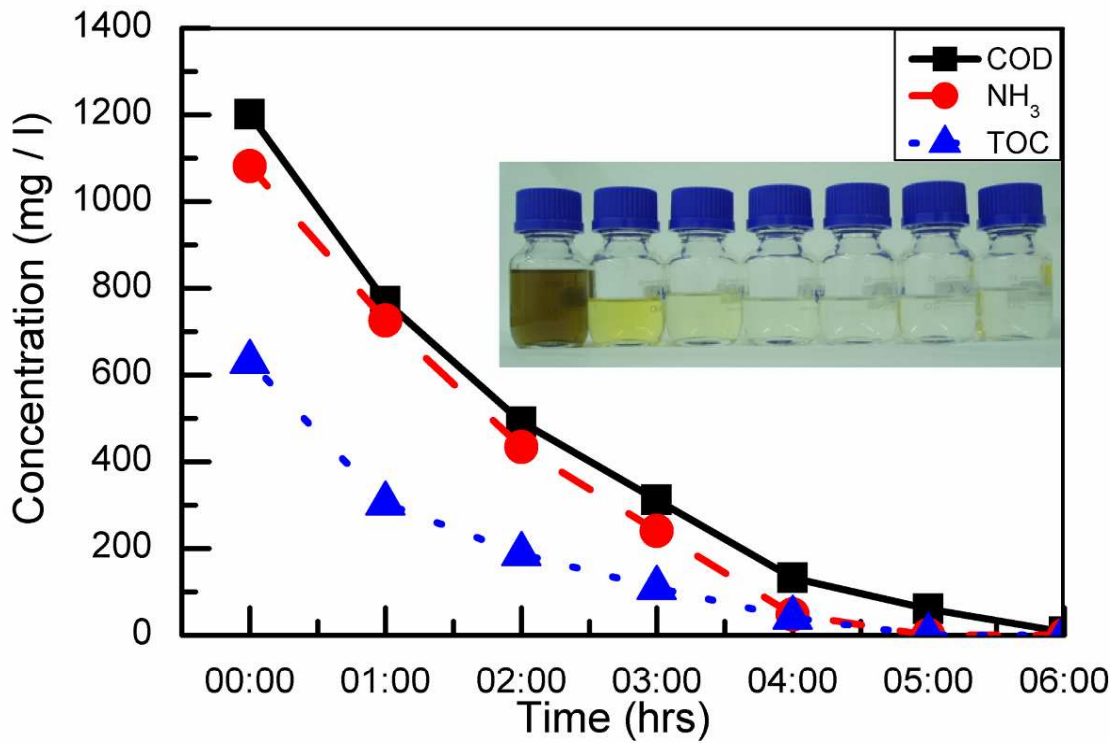

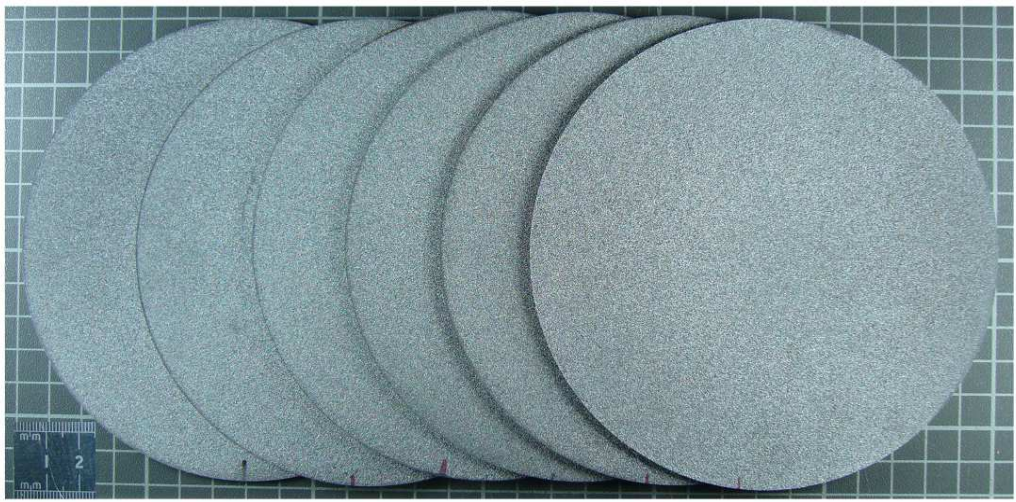

**Figure 16.** Top: Concentration of chemical oxygen demand (COD), ammonia and total organic carbon (TOC) as a function of time for landfill leachate treated with Diamox reactor. Inset image shows improved colour of the solution from before treatment and at one hour intervals during treatment.
Bottom: Typical boron doped diamond discs (138 mm diameter, thickness 0.5 to 1.3 mm) produced by Element Six for electrochemical applications.